\documentclass{emulateapj}
\usepackage{natbib}
\usepackage{times}
\usepackage{graphicx}
\usepackage{amsmath}
\usepackage{subfigure}

\newcommand{\kms}{km s$^{-1}$~}
\newcommand{\Msun}{M_\odot}
\newcommand{\moneynumber}{$398^{+178}_{-94}$ (98\% c.l.) }
\newcommand{\mainscen}{fiducial~}

\begin{document}

\title{Hundreds of Milky Way Satellites? Luminosity Bias in the Satellite Luminosity Function}
\shorttitle{Hundreds of Milky Way Satellites?}

\author{Erik J. Tollerud\altaffilmark{1}, James S. Bullock\altaffilmark{1}, Louis E. Strigari\altaffilmark{1}, Beth Willman\altaffilmark{2} }
\altaffiltext{1}{Center for Cosmology, Department of Physics and Astronomy, The University of California at Irvine, Irvine, CA, 92697, USA}
\altaffiltext{2}{Harvard-Smithsonian Center for Astrophysics, 60 Garden St., Cambridge, MA, 02138}

\begin{abstract}
We correct the observed Milky Way satellite luminosity function for luminosity bias using published completeness limits for the Sloan Digital Sky Survey DR5.   
Assuming that the  spatial distribution of Milky Way satellites tracks the subhalos found in the \emph{Via Lactea} $\Lambda$CDM N-body simulation, we show that there should be between $\sim 300$ and $\sim 600$ satellites within 400 kpc of the Sun that are brighter than 
the faintest known dwarf galaxies,
and that there may be as many as $\sim 1000$, depending on assumptions.   By taking into account completeness limits, we show that the radial distribution of known Milky Way dwarfs is consistent with our assumption that the full satellite population tracks that of subhalos.     These results alleviate the primary worries associated with the so-called missing satellites problem in CDM.  We show that future, deep wide-field surveys such as SkyMapper, the Dark Energy Survey (DES), PanSTARRS, and the Large Synoptic Survey Telescope (LSST) will deliver a complete census of ultra-faint dwarf satellites out to the Milky Way virial radius, offer new limits on the free-streaming scale of dark matter, and provide unprecedented constraints on the low-luminosity threshold of galaxy formation.
\end{abstract}

\section{Introduction}
\label{sec:intro}
It is well-established from simulations in the $\Lambda$CDM concordance cosmology that galaxy halos are formed by the merging of smaller halos \citep[e.g.][and references therein]{stewart08}.  As a result of their high formation redshifts, and correspondingly high densities, a large number of self-bound dark matter subhalos should survive the merging process and exist within the dark matter halos of $L_*$ galaxies like the Milky Way \citep{kl99ms,moo99ms,ZB03}.   A direct confirmation of this prediction has yet to occur.  Surveys of the halos of the Milky Way \citep[e.g.][]{will05wI,belfos06,walsh07boo,bel08LeoV} and Andromeda \citep[e.g.][]{martin06,majewski07,irwin08,mcconn08newands} have revealed only $\sim 20$ luminous dwarf satellites around each galaxy, approximately 1 order of magnitude fewer than the expected number of subhalos 
that are thought to be massive enough to form stars
\citep{kl99ms, moo99ms,die07ms,str07redef}.  
Of course, the mismatch is much more extreme (larger than a factor of $\sim 10^{10}$) 
when compared to the full mass function of CDM subhalos,  which is expected to rise as $\sim 1/M$ 
down to the small-scale clustering cutoff for the dark matter particle,  $M_{cut} \ll 1 M_\odot$ \citep
{Hofmann:2001bi,bertschinger06,profumo06,jk08}

 Astrophysical solutions to this missing satellites problem (MSP) include a reduction in the ability of small halos to accrete gas after reionization \citep{BKW,somer02reion,ben02reion}, and tidal stripping, which shrinks the mass of halos \emph{after} they have formed stars \citep{KR04}.      Although there are some promising new techniques
 in development for studying the formation of low-mass galaxies within a CDM context
 \citep{Robertson_Kravtsov,kauf07,orban08dsfh}, current models have trouble explaining many details, particularly regarding the lowest luminosity dwarfs (see below).  Other more exotic solutions rely on dark matter particles that are not cold \citep{HD00,K05,Cembranos:2005us,SKB} or non-standard models of inflation \citep{ZB03}, which produce small-scale power-spectrum cutoffs at $M_{cut} \sim 10^6-10^8 \Msun$, well above the cutoff scale in standard $\Lambda$CDM.    In principle, each of these scenarios leaves its mark on the properties of satellite galaxies, although multiple scenarios fit the current data. 
Hence, precisely determining the shape and normalization of the lowest end of the luminosity function is critical  to understanding how faint galaxies form and how the efficiency of star formation is suppressed in the smallest dark matter halos.  

The Sloan Digital Sky Survey (SDSS; \citealt{SDSSDR5}) has revolutionized our view of the Milky Way and its environment, doubling the number of known dwarf spheroidal (dSph) galaxies over the last several years  \citep[e.g.][]{will05wI,z06,z06b,belfos06,grillmair06,walsh07boo,Irwin07}.  Many of these new dSph galaxies are ultra-faint, with luminosities as low as $\sim 1000 L_\odot$, faint enough to evade detection in surveys with limits sufficient for detecting most previously known Local Group dwarfs \citep{whiting07photolims}.  In addition to providing fainter detections, the homogeneous form of the SDSS allows for a much better understanding of the statistics of detection.  Unfortunately, given the inherent faintness of the newly discovered dSph's and the magnitude-limited nature of SDSS for such objects, a derivation of the full luminosity function of satellites within the Milky Way halo must include a substantial correction for more distant undetectable satellites.  \citet{kop07} provided an important step in this process by performing simulations in order to quantify the detection limits of the SDSS and estimated the luminosity function by applying these limits to some simple radial distribution functions, finding $\sim 70$ satellites and a satellite luminosity function consistent with a single power law of the form $dN/dL_V \sim L^{-1.25}$.  

Our aim is to take the detection limits for SDSS Data Release 5 (DR5) as determined by \citet{kop07} and combine them with a CDM-motivated satellite distribution. In order to provide a theoretically-motivated estimate for the total number of Milky Way satellite galaxies, 
we adopt the distributions of subhalos in the \emph{Via Lactea} simulation \citep{die06vl}, and assume these to be hosts of satellite galaxies. 
Note that for the purposes of this paper, we define a  ``galaxy'' as a stellar system that is embedded in a dark matter halo.

The organization of this paper is as follows: In \S \ref{sec:app} we describe the overall strategy and sources of data used for this correction, as well as discussing the validity of the assumptions underlying these data. \S \ref{sec:corr} describes in detail how the correction is performed and presents the resulting luminosity functions for a number of possible scenarios, while \S \ref{sec:future} discusses the prospects for detecting the as-yet unseen satellites that the correction predicts.  \S \ref{sec:disc} discusses some of the caveats associated with the technique used for this correction, as well as the cosmological implications of the presence of this many satellite galaxies.  In \S \ref{sec:conc} we draw some final conclusions.

\section{Approach \& Data Sources}
\label{sec:app}
\subsection{Strategy}
The luminosity bias within the SDSS survey can be approximated by a characteristic heliocentric completeness radius, $R_{\rm comp}(M_{\rm V})$, beyond which a dwarf galaxy of absolute magnitude $M_{\rm V}$ cannot be observed.  This radial incompleteness is accompanied by a more obvious angular incompleteness -- the SDSS DR5 covers only a fraction $f_{\rm DR5} = 0.194$ of the sky, or a solid angle $\Omega_{\rm DR5} = 8000$ square degrees.
For the corrections presented in \S 3, we assume that magnitude $M_{\rm V}$ satellites with helio-centric distances $r > R_{\rm comp}(M_{\rm V})$ have not been observed, while satellites with $r \le R_{\rm comp}(M_{\rm V})$  have been observed, provided that are situated within the area of the sky covered by the SDSS footprint.       Given an observed number of satellites brighter than $M_{\rm V}$ within the survey area $\Omega_{\rm DR5}$ and within a radius $r = R_{\rm comp}(M_{\rm V})$, we aim to determine a correction factor, $c$, that gives the total number of satellites brighter than $M_{\rm V}$ within a spherical volume associated with the larger outer radius $R_{\rm outer}$: $N_{\rm tot} = c(r,\Omega) \, N_{\rm obs}(<r, <\Omega)$.   For our main results we count galaxies within $R_{\rm outer} = 417$ kpc, corresponding to the distance to the outermost Milky Way satellite (Leo T), although we do consider other $R_{\rm outer}$ values in \S \ref{sec:corr}.

It is useful (although not precisely correct) to think of the correction factor as a multiplicative combination of a radial correction factor and an angular correction factor, $c = c_{r} c_{\Omega}$.  The first correction, $c_r$, will depend on the radial distribution of satellites.    If there is no systematic angular bias in the satellite distribution, and if the experiment is performed many times, then the second correction factor should have an average value  $<c_{\Omega}> = 1/f_{\rm DR5} = 5.15$.  However, if the satellite distribution is anisotropic on the sky, the value of $c_\Omega$ for any particular survey pointing may be significantly different from the average.  An estimate of this anisotropy is essential in any attempt to provide a correction with meaningful errors.   Below, we use the satellite halo distribution in \emph{Via Lactea} to provide an estimate for the overall correction (see \S \ref{sec:vl}).  Note that in our final corrections presented in \S 3, we do not force the radial and angular corrections to be separable, but rather use a series of mock survey pointings within the simulation to calculate an effective correction, $c = c(r,\Omega)$, satellite by satellite.

 \begin{figure}[t!]
  \plotone{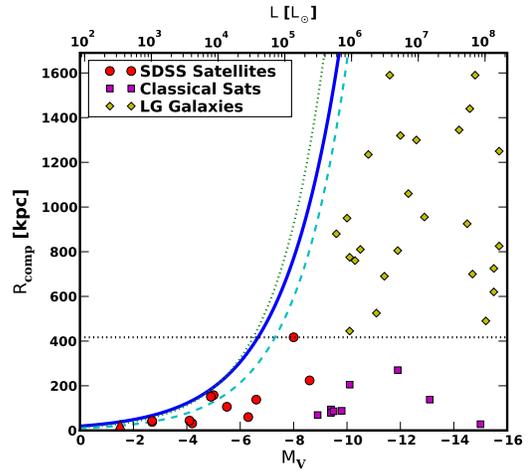}
 \caption{The completeness radius for dwarf satellites. The three rising lines show the helio-centric distance, $R_{\rm{comp}}$, out to which dwarf satellites of a given absolute magnitude are complete within the SDSS DR5 survey.  The solid is for the published detection limits in \citet{kop07}  and the other lines are the other two detection limits described in \S \ref{sec:sdss}.  
 The dotted black line at 417 kpc corresponds to our \mainscen adopted outer edge of the Milky Way halo satellite population ($R_{\rm outer}$).  The data points are observed satellites of the Milky Way and the Local Group.  The red circles are the SDSS-detected satellites, the only galaxies for which the detection limits actually apply, although the detection limits nevertheless also delineate the detection zone for more distant Local Group  galaxies (yellow diamonds).  Purple squares indicate classical Milky Way satellites.  The faintest object (red triangle) is Segue 1, which is outside the DR5 footprint.  }
 \label{fig:rvsmv}
 \end{figure}

\subsection{SDSS Detection Limits}
\label{sec:sdss}

\citet{kop07} constructed an automated pipeline to extract the locations of Milky Way satellites from the DR5 stellar catalog. They then constructed artificial dSph galaxy stellar populations (assuming a Plummer distribution), added them to the catalog, and ran their pipeline.  The detections of galaxies were used to construct detection thresholds as a function of distance  \citep[][Figure 8]{kop07}.  These thresholds are mostly constant as a function of surface brightness for $\mu \lesssim 30$ and linear with the logarithm of distance.  Hence, their detection threshold is well approximated as a log-linear relationship between the absolute magnitude of a dSph galaxy, $M_{\rm V}$, and the volume, $V_{\rm comp}(M_{\rm V})$,  out to which the DR5 could detect it.  Specifically, Figure 13 of \cite{kop07} implies
\begin{equation}
\label{eqn:complim}
V_{\rm comp} = 10^{( - a M_{\rm V} - b)} \, {\rm Mpc}^{-3},
\end{equation}
such that the completeness volume follows $V_{\rm comp} \propto L_{\rm V}^{5 a / 2}$. We adopt this form as the dwarf galaxy detection limit of DR5. 
  This volume may be related to a spherical completeness radius, $R_{\rm comp}$,  beyond which a dSph of a particular magnitude will go undetected:
\begin{equation}
R_{\rm comp}(M_{\rm V}) = \left(\frac{3}{4 \pi f_{DR5}}\right)^{1/3} 10^{(-a M_{\rm V}-b)/3} {\rm Mpc},
\label{eqn:detrad}
\end{equation}
where  $f_{\rm DR5} = 0.194$ is the fraction of the sky covered by DR5. 
For our \mainscen relation, we use the result presented in \citet[][Figure 13]{kop07}, which is fit by $a=0.60$ and $b=5.23$.  We note that for this value of $a$, we have $R_{\rm comp} \approx 66\:{\rm kpc} (L/1000 L_{\odot})^{1/2}$, and the SDSS is complete down to a fixed \emph{apparent} luminosity. The implied relationship between galaxy luminosity and corresponding helio-centric completeness radius is shown by the solid line in Figure \ref{fig:rvsmv}.  We also consider two alternative possibilities.   One is obtained by fitting the data in \citet{kop07} Table 3, which gives $a=0.684$ and $b=5.667$ (black dotted), and the other is a line that passes through the new SDSS satellites, on the assumption that some of them are of marginal detectability ($a=0.666$ and $b=6.10$; cyan dashed). 

It is important to recognize that, in principle, the detectability of satellites at a particular radius is not a step-function between detection and non-detection.  It also should not be spherically symmetric (independent of latitude with respect to the disk plane), nor should it be independent of other variables (such as satellite color or background galaxy density). However, \cite{kop07} found that a simple radial dependence provided a good description of their simulation results, and we adopt it for our corrections here.  They did, however, find that galaxies within this ``completeness'' boundary were not always detected at 100\% efficiency (depending on their distance and luminosity) and published detection efficiencies for the known SDSS dwarfs (their Table 3). We use these published detection efficiencies in our \mainscen corrections to the luminosity function below.  We also investigate how our results change when we assume 100\% efficiency in the correction.

 For reference, the horizontal dotted line in Figure \ref{fig:rvsmv} marks our \mainscen adopted $R_{\rm outer}$ radius for the Milky Way halo (slightly larger than the virial radius, in order to include Leo T).  According to this estimate, only satellites brighter than $M_{\rm V} \simeq -7$ are observable out to this radius.  The fact that the faintest dwarf satellite galaxies known are more than $4$ magnitudes fainter than this limit (see Table 1) immediately suggests that there are many more faint satellite galaxies yet to be discovered.

 \begin{figure}[t!]
 \plotone{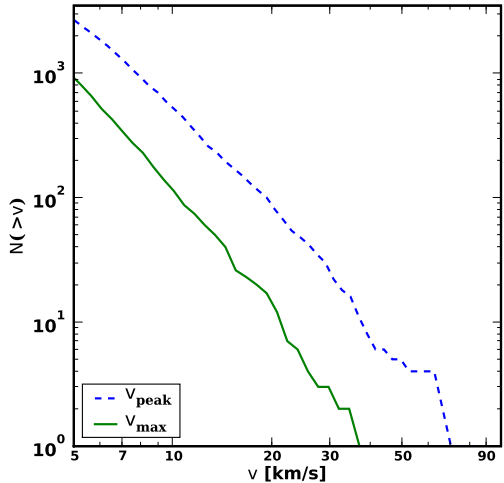}
 \caption{The cumulative count of \emph{Via Lactea} subhalos as a function of the (current) maximum circular velocity of the subhalo ($v_{\rm max}$, solid), and as a function of the largest maximum circular velocity ever obtained by the subhalo ($v_{\rm{peak}}$, dashed).   }
 \label{fig:nvsvmax}
 \end{figure}

\subsection{Via Lactea}
\label{sec:vl}
\citet*{die06vl} describe the \emph{Via Lactea} simulation, which is among the highest resolution $\Lambda$CDM N-body simulations of a Milky Way-like dark matter halo yet published.  The mass of Via Lactea is  $M_{200} \simeq 1.8 \times 10^{12}$ M$_{\odot}$, with a corresponding virial radius $R_{200} = 389$ kpc, where $M_{200}$ and $R_{200}$ are defined by the volume that contains 200 times the mean matter density. It resolves a large amount of substructure, recording 6553 subhalos with peak circular velocities  larger than $5$ \kms at some point in their history, of which 2686 are within our adopted $R_{\rm outer}=417$ kpc at $z=0$.    We use the public data release kindly provided by \citet{die06vl} in what follows~\footnote{\url{http://www.ucolick.org/\~{}diemand/vl/data.html}}.  

Figure \ref{fig:nvsvmax} presents the cumulative maximum circular velocity function, $N(>v_{\rm max})$, for \emph{Via Lactea} subhalos with halo-centric radius $R < 417$ kpc at $z=0$ (solid), along with the cumulative ``peak'' circular velocity function (dashed) for the same halos $N(>v_{\rm peak})$.  Here $v_{\rm peak}$ is the maximum circular velocity that the subhalos {\em ever} had over their history.   As emphasized by \citet{KR04} many subhalos have lost considerable mass over their history, and therefore $v_{\rm peak}$ may be a more reasonable variable to associate with satellite visibility than the (current) subhalo $v_{\rm max}$.
The information contained within this figure is presented elsewhere in the literature \citep{die06vl}, 
although we include it here for the sake of completeness.

\begin{figure}[t!]
\plotone{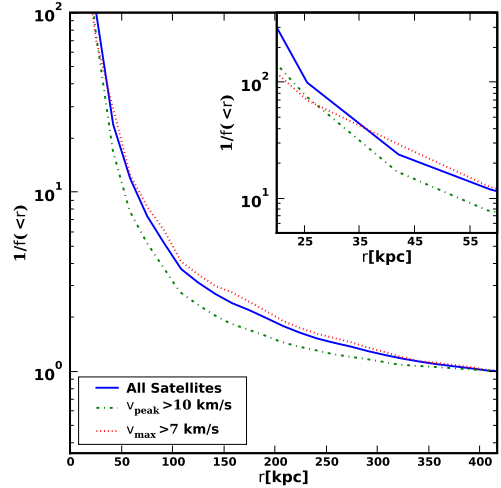}
\caption{Inverse of cumulative subhalo counts within the specified radius.  Here $f(<r)$ is the fraction of subhalos that exist within a given radius, normalized to unity at $R_{\rm outer} = 417$ kpc.  The radius is ``heliocentric", defined relative to  the (8,0,0) kpc position of the \emph{Via Lactea} simulation.   We include three populations of subhalos, as identified in the key. The inset focuses on the radial range where most of the new ultra-faint SDSS satellites have been detected.}
\label{fig:nvsr}
\end{figure}

An important ingredient in the luminosity bias correction is the assumed 
underlying radial distribution of satellites. We determine this 
distribution directly from~\emph{Via Lactea}.  
We are interested in determining the total number of satellites, $N_{\rm tot}$, 
given an observed number within a radius,  $N_{\rm obs}(<r)$. The correction 
will depend on the cumulative fraction of objects within a radius $r$ compared to the total count within some \mainscen outer radius: $f(<r)=N(<r)/N_{\rm tot}$.
The associated radial correction factor is then the inverse of the cumulative fraction, $c_r = f^{-1}(<r)$, such that
$N_{\rm tot} = c_r N_{\rm obs}(<r)$.    This correction factor  is shown in Figure \ref{fig:nvsr} for three different choices of subhalo populations: $v_{\rm peak} > 10$ \kms, $v_{\rm max} > 7$ \kms, and $v_{\rm peak} > 5$ \kms ( {\em i.e.} all subhalos in the \emph{Via Lactea} catalog). 

In order to mimic a heliocentric radial distribution,  in Figure \ref{fig:nvsr} we have placed the observer at a radius of $8$ kpc within a fictional disk centered on \emph{Via Lactea}.  The orientation of the ``disk'' was chosen to be in the \emph{Via Lactea} $xy$ plane for all figures that show subhalo distributions, but we allow for a range of disk orientations for the full correction presented in \S \ref{sec:corr}.   We do this for completeness, but stress that our results are extremely insensitive to the choice of solar location, and are virtually identical if we simply adopt a vantage point from the center of \emph{Via Lactea}.  The total count is defined within our \mainscen $R_{\rm{outer}}$,  such that $f(<417$ kpc$) = 1.0$. An important result of this is that our correction does \emph{not} depend on the number of subhalos in Via Lactea, only on the shape of the distribution.  Still, the shape varies among some sub-populations of subhalo.  As noted in~\citet{KR04},~\citet{die06vl} and~\citet{madau081e8ell}, subhalos chosen to have a large  $v_{\rm peak}$ tend to be more centrally concentrated.  However, 
the correction we apply is fairly insensitive to the differences between these choices of subhalo populations. 

The most important corrections will be those for the faintest galaxies, which are just observable at local distances $r \lesssim 50$ kpc.     It is thus clear  that the radial correction factors associated with the three different subhalo populations show very little differences at the radii of relevance (see inset in Figure \ref{fig:nvsr}).    Motivated by this result, we use the full \emph{Via Lactea} subhalo catalog in our \mainscen model because it provides a larger statistical sample of subhalos.  However, for the sake of completeness, we present the final corrected counts for our corrections using the  $v_{\rm peak} >  10$ \kms and $v_{\rm max} > 7$ \kms samples in Table \ref{tab:res} below.  

\begin{figure}[t!]
\begin{center}
\plotone{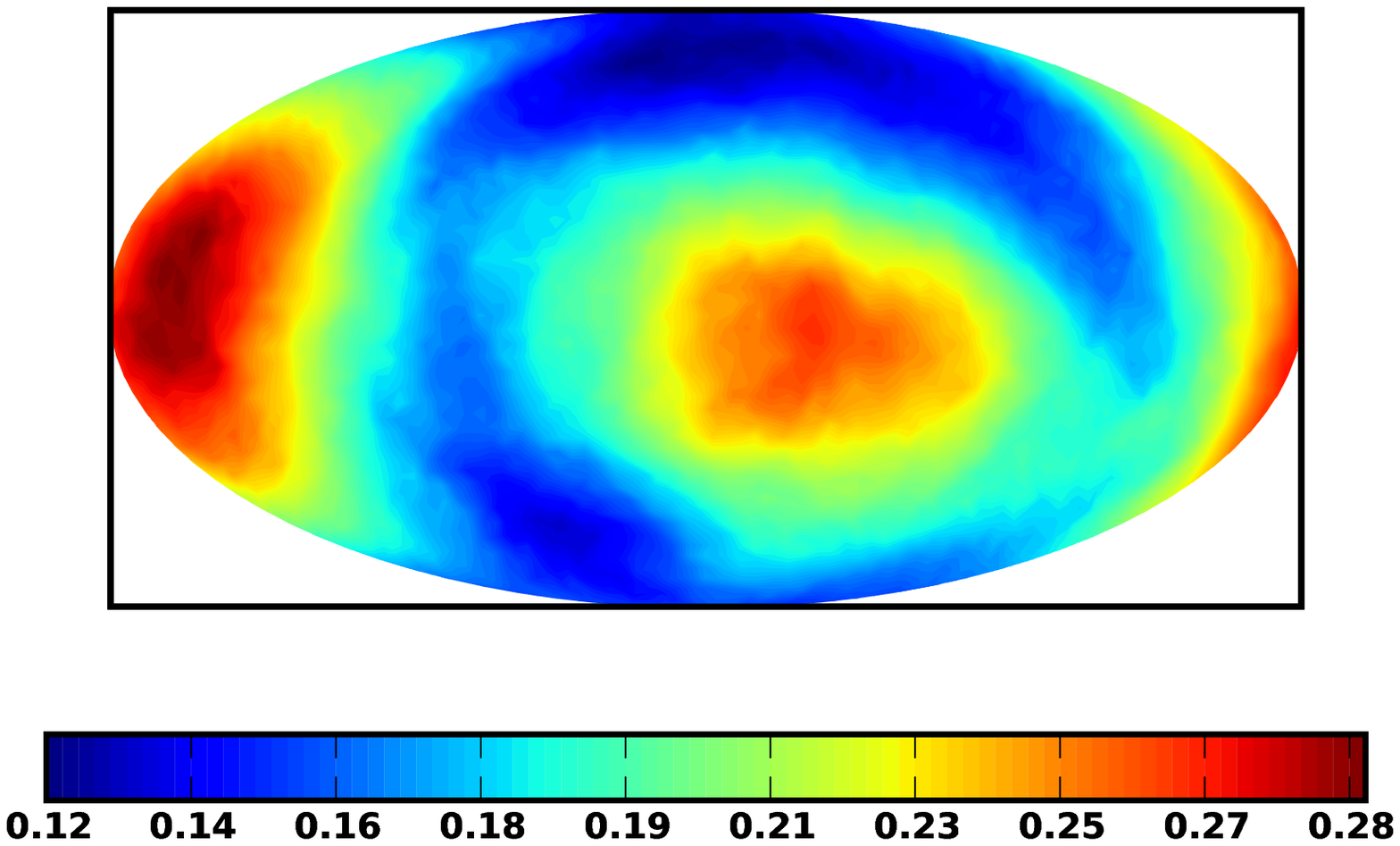}
\plotone{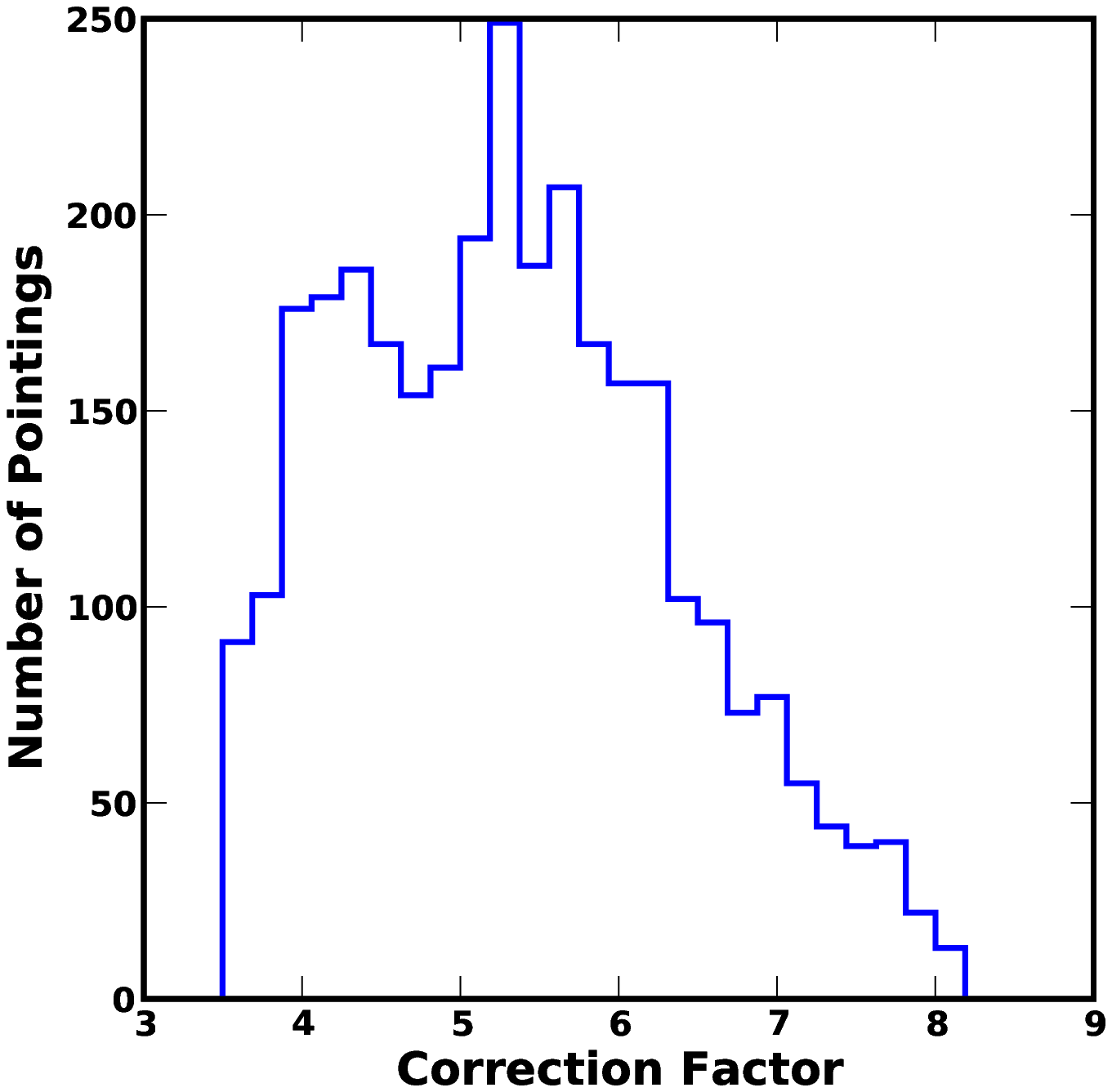}
\end{center}
\caption{Example of subhalo angular anisotropy.  Upper: Hammer projection map of the angular anisotropy in \emph{Via Lactea} subhalos with $v_{\rm peak} > 5$ \kms. Colors at each point indicate the fraction of subhalos, $f(<\Omega)$, within $417$ kpc that are contained within an angular cone of $\Omega_{\rm DR5} =8000$ square degrees, in order to match the area of SDSS DR5.  Lower: Distribution of the angular correction factors, $c_\Omega = 1/f(<\Omega)$,  which must be applied to the count within 8000 square degrees in order to return the full number of subhalos within $417$ kpc.   The histogram includes 1000 pointings, evenly spaced over the sky.    For reference, the sky coverage of DR5 is $f_{DR5} = 0.194$, which implies an angular correction factor of $1/f_{DR5} = 5.15$.   As it must, $c=5.15$ matches the mean of the distribution, but not the median, which is $c = 5.27$.  Note that these correction factors do not yet include the effects of radial incompleteness/luminosity bias.}
\label{fig:aniso}
\end{figure}

\emph{Via Lactea} can also be used to extract the angular distribution of subhalos.  This will be important for determining the errors on the overall satellite abundance that may arise from limited sky coverage.  The SDSS DR5 footprint covers 8000 square degrees, or a fraction $f_{DR5} = 0.194$ of the sky.  While this is a sizable fraction of the sky, the upper image in Figure \ref{fig:aniso} shows that the subhalo distribution (projected out to the virial radius, in this case) is quite anisotropic even on this scale.   The color of each pixel is set by the fraction of all the subhalos, $f(<\Omega)$, that exist within an area of $\Omega_{\rm DR5} = 8000$ square degrees, centered on this pixel.   We see clearly that the fraction of satellites within a DR5-size region can vary considerably  (from $0.12$ to $0.28$)  depending on the pointing orientation.   The same information is shown as a histogram in the lower panel, now presented as the implied multiplicative ``sky coverage'' correction factor, $c_{\Omega} = 1/f(<\Omega)$.  Each of the 3096 pixels in the sky map is included as a single count in the histogram.   Note that while the average correction factor is $c_{\Omega} = 1/f_{DR5} = 5.15$ (as expected) the correction can vary from just $\sim 3.5$ up to $\sim 8.3$ depending on the mock survey's orientation.

Before moving on, we note that the expected anisotropy may contain a possible solution to the ``missing inner satellites problem'' described in \citet{die06vl}, which notes that there are $\sim 20$ subhalos in \emph{Via Lactea} within the innermost parts of the halo, while only a few galaxies are actually observed.  While numerical effects may become important for these subhalos given their sizes and how close they are to the center of the host, it may also simply be an observational coverage effect.  If only subhalos within 23 kpc (the distance to Segue 1) are considered, $\sim 15\%$ of the sky has one or zero subhalos visible in a survey the size of SDSS.  This means that even if all of the subhalos host galaxies, there is a 15\% chance that we will at most see a galaxy like Segue 1.  The fact that Segue 1 was only recently discovered shows that there is easily room for a significant number of inner satellites that have gone as yet undetected simply because a survey like SDSS is necessary to uncover such objects, but over the entire sky instead of just a fifth.

\begin{deluxetable*}{cccccc}
\tablecolumns{6}
\tablecaption{Properties of known Milky Way Satellite galaxies. Data are from \citet{both88,mateo98rev,grebel03dsphprog,sandg07,martin08,dejong08leot}. }
\tablehead{
  \colhead{Satellite} &
  \colhead{M$_V$} &
  \colhead{L$_V$[$L_{\odot}$]} &
  \colhead{$d_{\rm sun}$[kpc]} &
  \colhead{$R_{\rm half}$[pc] \tablenotemark{a}} &
  \colhead{$\epsilon$ \tablenotemark{b}}
}
\startdata
\cutinhead{SDSS-discovered Satellites}
\tablenotemark{$\star$}Bo{\"o}tes I & -6.3 & $2.83\times10^4$ & 60 & 242 & 1.0\\
\tablenotemark{$\star$}Bo{\"o}tes II & -2.7 & $1.03\times10^3$ & 43 & 72 & 0.2\\
\tablenotemark{$\star$}Canes Venatici I & -8.6 & $2.36\times10^5$ & 224 & 565 & 0.99\\
\tablenotemark{$\star$}Canes Venatici II & -4.9 & $7.80\times10^3$ & 151 & 74& 0.47\\
\tablenotemark{$\star$}Coma & -4.1 & $3.73\times10^3$ & 44 & 77& 0.97\\ 
\tablenotemark{$\star$}Hercules & -6.6 & $3.73\times10^4$ & 138 & 330& 0.72\\ 
\tablenotemark{$\star$}Leo IV & -5.0 & $8.55\times10^3$ & 158 & 116 & 0.79\\ 
\tablenotemark{$\star$}Leo T & -8.0 & $5.92\times10^4$ & 417 & 170 & 0.76\\ 
\tablenotemark{$\dagger$}Segue 1 & -1.5 & $3.40\times10^2$ & 23 & 29 & 1.0\\
\tablenotemark{$\star$}Ursa Major I & -5.5 & $1.36\times10^4$ & 106 & 318 & 0.56\\ 
\tablenotemark{$\star$}Ursa Major II & -4.2 & $4.09\times10^3$ & 32 & 140 & 0.78\\ 
\tablenotemark{$\star$}Willman 1 & -2.7 & $1.03\times10^3$ & 38 & 25 & 0.99\\ 

\cutinhead{Classical (Pre-SDSS) Satellites}

Carina & -9.4 & $4.92\times10^5$ & 94 & 210 & -\\ 
\tablenotemark{$\star$}Draco & -9.4 & $4.92\times10^5$ & 79 & 180 & 1.0\\ 
Fornax & -13.1 & $1.49\times10^7$ & 138 & 460 & -\\ 
LMC & -18.5 & $2.15\times10^9$ & 49 & 2591 & -\\ 
Leo I & -11.9 & $4.92\times10^6$ & 270 & 215 & 1.0\\ 
\tablenotemark{$\star$} Leo II& -10.1 & $9.38\times10^5$ & 205 & 160 & 1.0\\ 
Ursa Minor & -8.9 & $1.49\times10^5$ & 69 & 200 & -\\ 
SMC & -17.1 & $5.92\times10^8$ & 63 & 1088 & -\\ 
Sculptor & -9.8 & $7.11\times10^5$ & 88 & 110 & -\\ 
Sextans & -9.5 & $5.40\times10^5$ & 86 & 335 & -\\ 
Sagittarius & -15 & $8.55\times10^7$ & 28 & 125 & -\\ 
\enddata

\tablenotetext{a}{Satellite projected half light radius.}
\tablenotetext{b}{Detection efficiency from \cite{kop07}.}
\tablenotetext{$\star$}{Galaxy sits within the SDSS DR5 footprint.}
\tablenotetext{$\dagger$}{Satellite is not used in \mainscen LF correction.}

\label{tab:sats}
\end{deluxetable*}

\subsection{Observed Satellite Galaxies}
\label{sec:obssats}
A variety of authors have identified dSph's in the DR5 footprint, including some that straddle the traditional boundaries between globular clusters and dSph's.  For example, Willman 1  was originally of unclear classification~\citep{will05wI}, but is now generally recognized as being a dark matter dominated dSph~\citep{mar07will1,strigari08}.  Table \ref{tab:sats} lists the complete set of dwarf galaxies from DR5 used in this analysis. For the SDSS dwarfs, we use the luminosities from \citet{martin08}, which presents a homogeneous analysis, for all of the SDSS dwarfs, as well as \citet{dejong08leot} for Leo T, which is based on much deeper photometry.  We note that these results are in generally good agreement with pre-existing analyses \citep{will05wI,belfos06,sandg07,walsh07boo,walsh07boo2}. Here, we also list the ``classical'' dwarfs, which were discovered before the SDSS, along with Segue 1, which was discovered from data for the SDSS-II SEGUE survey \citep{bel07catsdogs}.  All of the objects we list in this table have large mass-to-light ratios~\citep{mar07will1,sandg07,strigari08}.   

For our \mainscen corrections, following the convention of \citet{kop07}, we have not included Segue 1, as it does not lie inside the DR5 footprint and hence the published DR5 detection limits are not applicable.  We do include Segue 1 in an alternative correction scenario below (see Table \ref{tab:res}).  We do not correct  the classical dwarf satellite galaxies for luminosity bias or sky coverage, because appropriate detection limits for these classical dwarf satellites are unclear given that they are not part of a homogeneous survey like SDSS.  We assume that all satellites within those magnitude bins would have been discovered anywhere in the sky, with the possible exception of objects at low Galactic latitudes, where Milky Way extinction and contamination become significant \citep{will04}.  This assumption is conservative in the sense that it will bias our total numerical estimate low, but it is only a minor effect, as our correction  described in \S \ref{sec:corr} is dominated by low luminosity satellites.  
 
\begin{figure}[t!]
\plotone{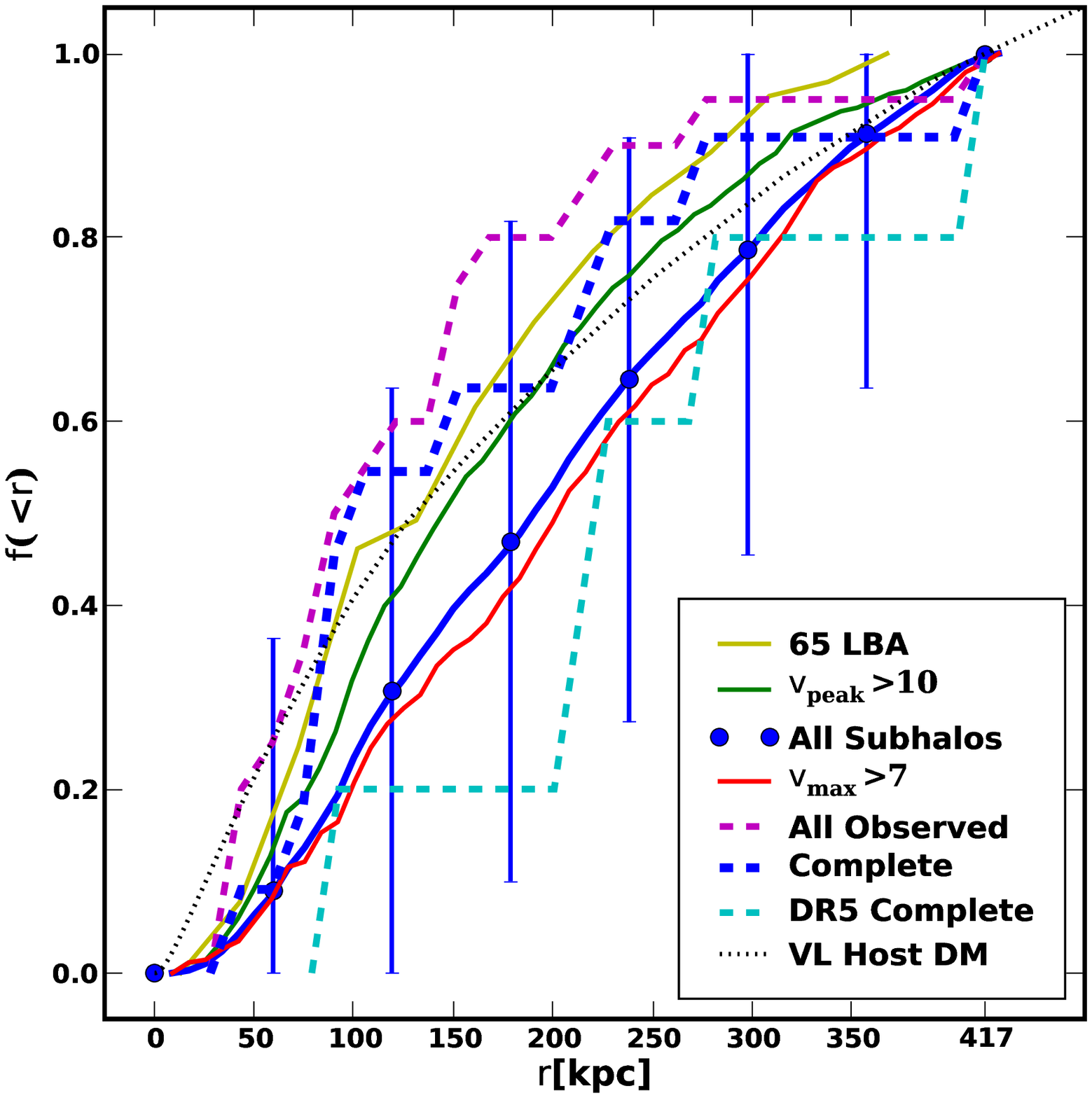}
\caption{Radial distributions for various populations of observed satellites as compared to several sample subhalo distributions in \emph{Via Lactea}.  Solid lines are four populations of \emph{Via Lactea} subhalos within 417 kpc: from top to bottom, 65 largest before accretion \citep[see][Figure 7]{madau081e8ell}, $v_{\rm peak} > 10$ \kms, All ({\it i.e.} $v_{\rm peak} > 5$ \kms), and $v_{\rm max}>7$ \kms.  The dashed lines are observed satellite distributions.  From top to bottom, we have the``All Observed''(magenta), which consists of all known Milky Way dSph satellites;  the ``Complete'' (blue) distribution corresponds to only those with magnitudes corresponding to $R_{\rm comp} \ge 417$ kpc; ``DR5 Complete'' (cyan) corresponds to satellites brighter than the completeness magnitude, and also in the DR5 footprint -- these are the only observationally complete and homogeneous sample.  The dotted line is the global dark matter distribution for the \emph{Via Lactea} host halo. Error bars (98 \% c.l. ) are derived by randomly sampling from the \emph{Via Lactea} subhalos the same number of satellites as in the ``Complete'' sample (11).  Note that this means the error bars only apply for comparing the solid lines to the dashed line, and that they are correlated, but still represent the scatter of individual bins in any possible observed sample of 11 satellites from {\emph Via Lactea}. }
\label{fig:raddist}
\end{figure}

Before we use the radial distribution of \emph{Via Lactea} subhalos to correct the observed luminosity function, it is important to investigate whether this assumption is even self consistent with the data we have on the radial distribution of known satellites.   The relevant comparison is shown in Figure \ref{fig:raddist}. 
We have normalized to an outer radius $R_{\rm outer} = 417$ kpc (slightly larger than the \emph{Via Lactea} virial radius) in order to allow a comparison that includes the DR5 dwarf Leo T; this extension is useful because the known dwarf satellite count is so low that even adding one satellite to the distribution increases the statistics significantly.

The radial distribution of all 23 known Milky Way satellites is shown by the magenta dashed line in Figure \ref{fig:raddist}.  The four solid lines show radial distributions for four choices of subhalo populations: the 65 largest $v_{\rm peak}$(upper) subhalos (65 LBA) as discussed in \citet{madau081e8ell}, $v_{\rm peak} > 10$ \kms (upper-middle), $v_{\rm peak} > 5$ \kms  (lower-middle), and $v_{\rm max} > 7$ \kms (lower).  
We note that the all-observed profile is clearly more centrally concentrated than any of the theoretical subhalo distributions.  However, as shown in Figure \ref{fig:rvsmv}, our limited ability to detect faint satellite galaxies almost certainly biases the observed satellite population to be more centrally concentrated than the full population.

If we include only the 11 satellites (excluding SMC and LMC) that are bright enough to be detected within $417$ kpc ($M_{\rm V} \la  -7$), we obtain the thick blue dashed line.  This distribution is significantly closer to all of the theoretical subhalo distributions, and matches quite well within $r \sim 50$ kpc, where the incompleteness correction to the luminosity function will matter most.  It is still more centrally concentrated then the distribution of all subhalos, however, as has been noted in the past \citep[at least for the classical satellites -- e.g.][]{will04rad,die04subhbias,KR04}.   In order to more rigorously determine whether the theoretical distribution is consistent with that of the 11 ``complete'' satellites, we have randomly determined the radial distribution of $v_{\rm peak} > 5$ \kms subhalos using 1000 subsamples of 11 subhalos each. The error bars reflect 98 \% c.l. ranges from these subsamples, although they are correlated and hence represent only a guide to the possible scatter about the points in the distribution. From this exercise, we can roughly conclude that the complete observed distribution is consistent with being a random subsample of the $v_{\rm peak} > 5$ \kms  population.  We also note that the $v_{\rm peak} >10$ \kms distribution fits even more closely, well within their 68\% distribution (errors on this thoretical distribution are not shown for figure clarity, but they are similar in magnitude to those on the central line).  

To further investigate the issue of consistency between the various radial distributions, we determine the KS-probability for various distributions. The results are shown in Table \ref{tab:radKS}, where  in each case the $p$-value gives the probability that the null hypothesis is correct (i.e. that the two distributions are drawn from the same underlying distribution). While the distribution of all satellites is clearly inconsistent with the subhalo distributions, choosing only the complete satellites improves the situation, giving a reasonable probability of the complete distribution matching the $v_{\rm peak} >5$ \kms distribution, and an even better match for the distributions that are cut on $v_{\rm peak}$, as expected from the error bars.

Finally, we note that our definition of ``complete'' is conservative, since no satellites fainter than $M_{\rm V} = -8.8$ were known before the SDSS survey.  If we include only the (five) satellites bright enough to be detectable to $R_{\rm outer}$ that exist within the well understood area of the DR5 footprint, we obtain the radial profile shown by the dashed cyan line (``DR5 Complete''), which is even less centrally concentrated than any of the theoretical lines.  While this distribution is marginally consistent with the other distributions based on the KS-test $p$-values, caution is noted given the small number of data points (5).  We conclude that while more data (from deeper and wider surveys) is absolutely necessary in order to determine the radial distribution of Milky Way satellites and to compare it with theoretical models, the assumption that the underlying distribution of satellites tracks that predicted for subhalos is currently consistent with the data.  We adopt this assumption for our corrections below -- specifically, we use the $v_{\rm peak} > 5$ \kms distribution for our \mainscen case.  While the $v_{\rm peak} > 10$ \kms distribution fits the observed satellites more closely, we show in \S \ref{sec:corr} that the various subhalo distributions affect only the final satellite counts by a factor of at most $\sim 1.4$.  We chose to use all \emph{Via Lactea} subhalos (corresponding to $v_{\rm peak} > 5$ \kms) as our \mainscen case in order to reduce statistical noise from the smaller number of satellites in the $v_{\rm peak} > 10$ population -- note that numerical effects will generally produce an under-correction (giving a smaller number of satellites), as described in \S \ref{sec:corr}.

Previous studies have pointed to discrepancies between simulations and observations of other galaxies' satellite distributions \citep[e.g.][]{die04subhbias,chen0radsdss}.  While this is of concern, much of these data are around clusters or other environments very different from that of the Local Group.  Furthermore, the detection prospects for faint satellites around the Milky Way are of course far better than can be achieved for other galaxies, and the region of the luminosity function that is of importance for this correction (e.g. $M_{\rm V}>-7$) has never been studied outside the Local Group.  
 
\begin{deluxetable}{ccc}[t!]
\tablecolumns{3}
\tablecaption{KS test results for radial distributions shown in Figure \ref{fig:raddist}.}
\tablehead{
  \colhead{Distribution 1} &
  \colhead{Distribution 2} &
  \colhead{$p$-value} 
}
\startdata
All Observed & 65 LBA & 57.6\% \\
All Observed & $v_{\rm peak} > 10$  & 10.3\% \\
All Observed & All Subhalos & 0.4\% \\
All Observed & $v_{\rm max} > 7$ & 0.1\% \\
Complete & 65 LBA & 95.1\% \\
Complete& $v_{\rm peak} > 10$ & 38.2\% \\
Complete& All Subhalos & 12.1\% \\
Complete & $v_{\rm max} > 7$ & 8.2\% \\
DR5 Complete & 65 LBA  & 8.6\% \\
DR5 Complete &  $v_{\rm peak} > 10$ & 14.4\% \\
DR5 Complete& All Subhalos & 48.2\% \\
DR5 Complete &  $v_{\rm max} > 7$ & 63.1\% \\
\hline
All Observed & Complete & 87.6\% \\
All Observed & DR5 Complete & 6.6\% \\
Complete & DR5 Complete & 41.1\%

\enddata
\label{tab:radKS}
\end{deluxetable}

\section{Luminosity Function Corrections}
\label{sec:corr}

\begin{figure}[t!]
\plotone{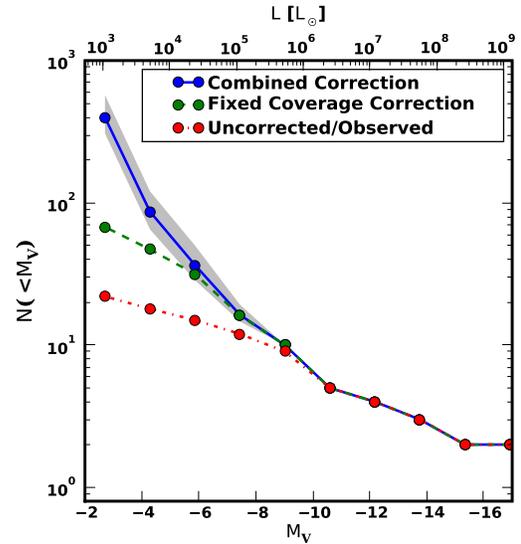}
\caption{Luminosity function as observed (lower), corrected for only SDSS sky coverage (middle), and with all corrections included (upper). Note that the classical (pre-SDSS) satellites are uncorrected, while new satellites have the correction applied.  The shaded error region corresponds to the 98\% spread over our mock observation realizations.
Segue 1 is not included in this correction.}
\label{fig:moneyplot}
\end{figure}

The cumulative number of known Milky Way satellites brighter than a given magnitude is shown by the lower (red) dashed line with solid circles in Figure \ref{fig:moneyplot}.  This luminosity function includes both the classical dwarf satellites and the fainter, more recently-discovered SDSS satellites (excluding Segue 1 for reasons discussed above).  By simply multiplying the SDSS satellite count by the inverse of the DR5 sky fraction ($1/f_{DR5}$) and adding this to the classical dwarf satellite count, we produce the (green) long-dashed line.  This first-order correction provides an extremely conservative lower estimate on the total Milky Way satellite count by ignoring the details of luminosity bias in the SDSS.

We use the following series of mock surveys of the \emph{Via Lactea} subhalo population in order to provide a more realistic correction.   First, an observer is positioned at distance of $8$ kpc  from the center of \emph{Via Lactea}.  We then define an angular point on the sky from this location and use it to center a mock survey of solid angle $\Omega_{\rm DR5} = 8000$ square degrees.  We allow this central survey position to vary over the full sky using 3096 pointings that evenly sample the sky.   Although we find that the absolute position of the observer does not affect our results significantly, we also allow the observer's position to vary over six specific locations, at  $(x,y,z) = (\pm 8, 0, 0), (0, \pm 8, 0)$, and $(0, 0, \pm8)$ on the \emph{Via Lactea} grid.  We acknowledge that there are (contradictory) claims in the literature concerning whether satellite galaxies are preferentially oriented (either parallel or perpendicular) with respect to galaxy disks  \citep{kr05satdisk,metz07satdist,kuhlen07shapes,wang08sdssalignment,falt08align}, and equally contradictory claims regarding how disks are oriented in halos \citep{zent05aniso,bal05diskalign,dutton07diskalgin}.  
If there were a preferential orientation, then the appropriate sky-coverage correction factors would need to be biased accordingly, but for our correction, we make no assumptions about the orientation of the ``disk''.  Therefore, any uncertainty in the correct orientation of the disk
 is contained within the errors we quote on our counts.
In the end, we produce 18576 equally-likely mock surveys each with their own correction factors, and use these to correct the Milky Way satellite luminosity function for angular and radial incompleteness.

For each of the mock surveys, we consider every DR5 satellite ($i = 1, ..., 11$, sometimes 12 ) with a helio-centric distance within $R_{\rm outer} = 417$ kpc  and determine the total number of objects of its luminosity that should be detectable.  Specifically, if satellite $i$ has a magnitude $M_{\rm V}^{i}$ that is too faint to be detected  at $R_{\rm outer}$ ({\it i.e.} if $M_{\rm V}^{i} \gtrsim -7$), then we determine the number of \emph{Via Lactea} subhalos, 
$N(r<R_{\rm comp},\Omega<\Omega_{\rm DR5}$), that are situated within an angular cone of size $\Omega_{\rm DR5}$ and within a helio-centric radius $R_{\rm comp}(M_{i})$.    We then divide the total number of {\em Via Lactea} subhalos, $N_{\rm tot}$ by this ``observed" count 
 and obtain a corrected estimate for the total number satellites of magnitude $M_{\rm V}^{i}$:
\begin{equation}
\label{eqn:corr}
c^i = \frac{N_{\rm tot}}{N(r<R_{\rm comp}(M_{\rm V}^i),\Omega < \Omega_{\rm DR5})}.
\end{equation}
If the SDSS satellite $i$ is bright enough to be seen at $R_{\rm outer}$, then $R_{\rm comp}$ is replaced by
$R_{\rm outer}$ in the above equation.  In this case, the correction factor accounts only for angular incompleteness.
This method allows us to produce distributions of correction factors for each $M_{\rm V}$ of relevance. Note also that because the correction factor is the fraction of subhalos in the cone, it depends only on the distributions of \emph{Via Lactea} subhalos, rather than directly depending on the total number, rendering it insensitive to the overall subhalo count in \emph{Via Lactea}.  Furthermore, any subhalo distributions that do not have enough subhalos to fully accommodate all possible pointings are simply given the correction factor $1/N_{\rm tot}$, meaning that numerical effects tend to \emph{under}correct, producing a conservative satellite estimate.

\begin{figure}[hbt]
\epsscale{0.82}
\plotone{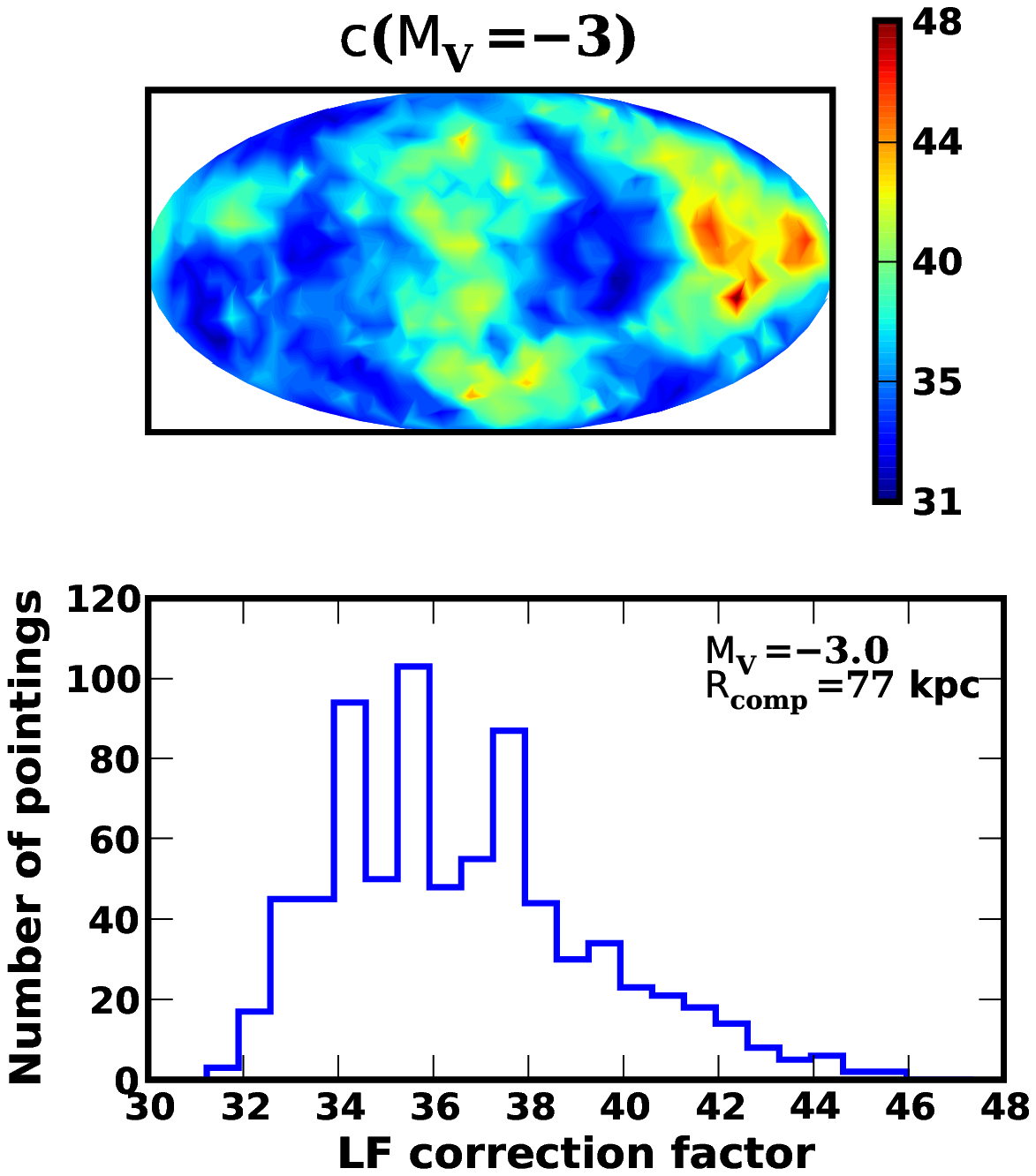}
\plotone{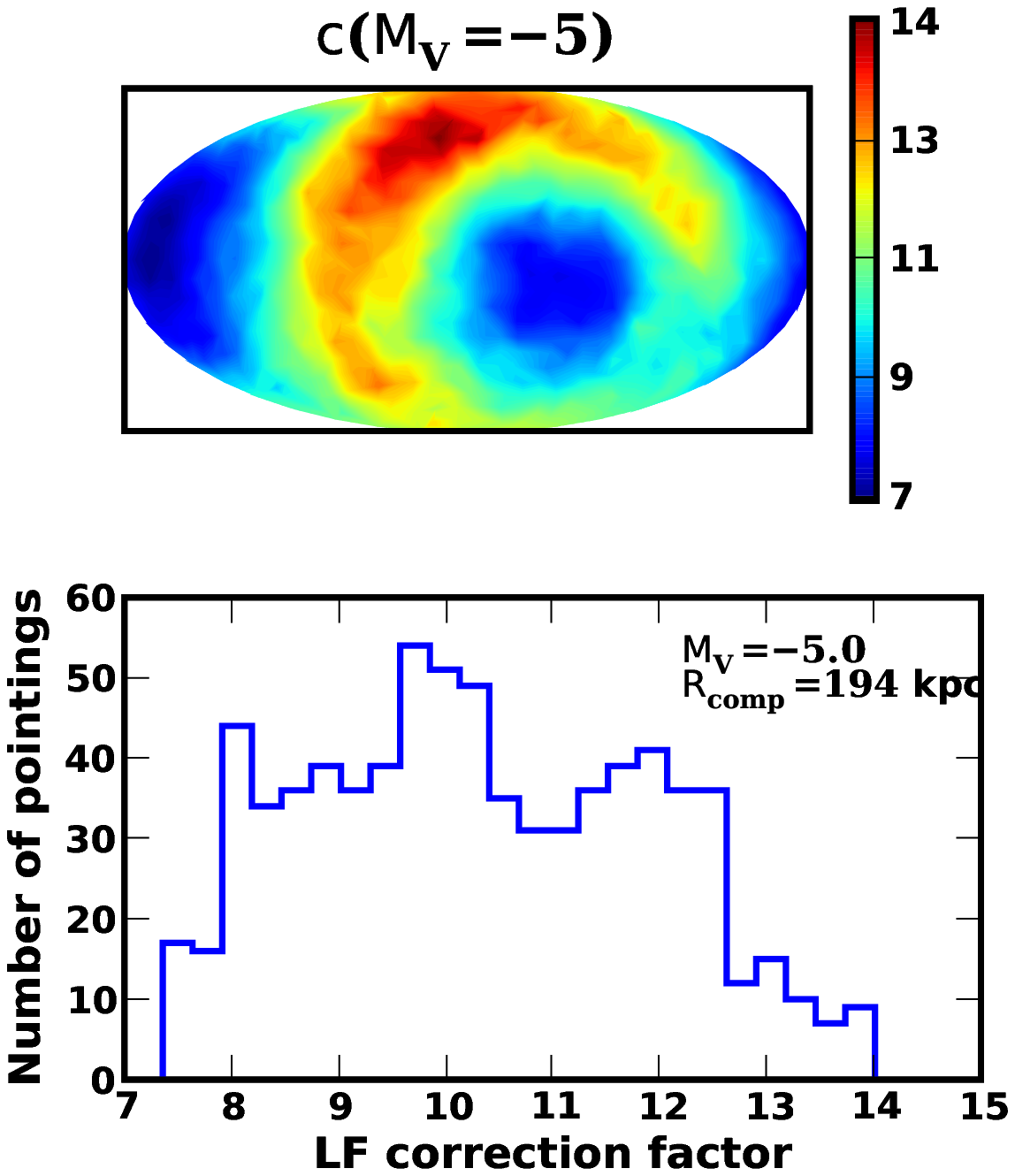}
\plotone{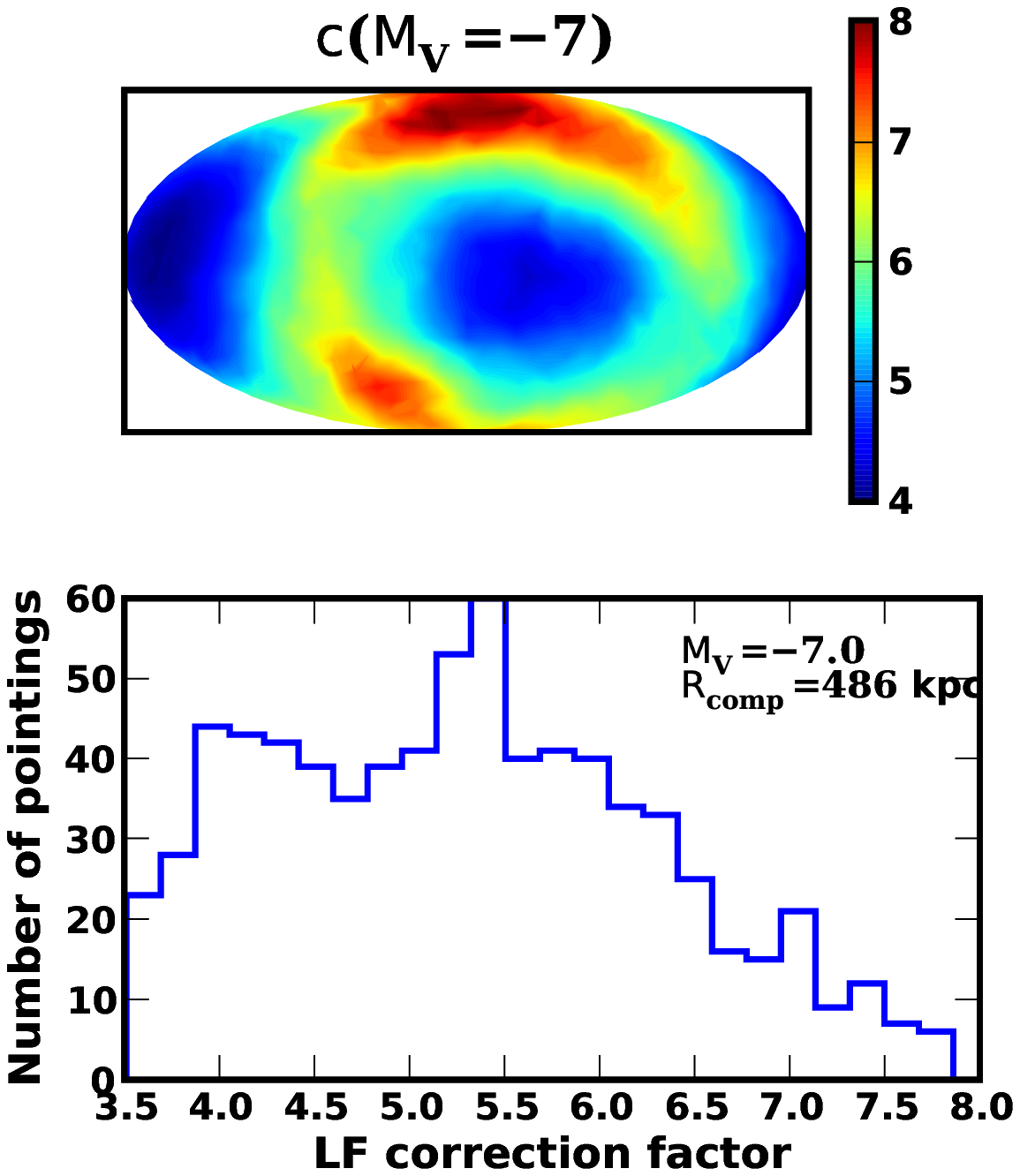}
\caption{Sky anisotropy maps and corresponding distributions for 3 sample limiting magnitudes.  The sky projections are the corresponding correction factor ({\it i.e.} inverse of the enclosed fraction of total satellites) for a DR5-sized survey pointing in a given direction corresponding to a satellite of a particular magnitude.  The histograms show the distribution of correction factors from evenly sampling the sky maps.  From top to bottom, limiting magnitudes are  $M_V=-3,-5,-7$ corresponding to $R_{\rm comp}=77,194,486$ kpc.}
\label{fig:3dists}
\end{figure}

Three example distributions of number-count correction factors are shown in Figure \ref{fig:3dists} for hypothetical objects of luminosity $M_{\rm V} = -3, -5, -7$ and corresponding completeness radii $R_{\rm comp} = 77,194, $ and $486$ kpc.  We see that while the brightest objects typically have correction factors of order the inverse of the sky coverage fraction, $\sim 5$, the faintest objects can be under-counted by a factor of $\sim 100$ or more. Note that to produce helio-centric sky coverage maps, we must assume a plane in \emph{Via Lactea} in which we consider the Sun to lie -- for Figure \ref{fig:3dists}, this is the $xy$ plane, but for our final corrections we consider all possible orientations (described above).

We construct corrected luminosity functions based on each of the 18576 mock surveys by generating a 
cumulative count of the observed satellites.  We weigh each satellite $i$
 by its associated correction factor $c^i$ and (in our \mainscen case) its detection efficiency $\epsilon^i$.  
For each of the new satellites we use the quoted detection efficiencies from  \citet[][Table 3]{kop07}. 
We reproduce those efficiencies in our Table 1, and we assume $\epsilon = 1$ for all of the classical 
satellites that are not within the DR5 footprint.   
Explicitly, the cumulative luminosity function for a given pointing is:
\begin{equation} 
n(<M_{\rm V})=\sum_{i}^{<M_{\rm V}} \frac{c(M_{\rm V}^{i})}{\epsilon^i}.
\label{eqn:histcorr} 
\end{equation}
For a given scenario (subhalo population, completeness limits, and detection efficiencies for each satellite)
we determine the luminosity function for each pointing and disk orientation.
We are then able to calculate a median luminosity function and scatter for each scenario.

Our \mainscen corrected luminosity function shown by the upper blue solid line in Figure \ref{fig:moneyplot}, and the shaded band spans the 49\% tails of the distribution. 
 Note that the errors vanish around $M_V=-9$ because all satellites brighter than that are ``classical" pre-SDSS satellites and are left completely uncorrected on the conservative assumption that any objects brighter than this would have been detected previously.
Our \mainscen scenario counts galaxies within a radius $R_{\rm outer} = 417$~kpc and
excludes  Segue 1 from the list of corrected satellites because it is not within the
DR5 footprint.   In addition, this scenario uses quoted detection efficiencies $\epsilon$ from \cite{kop07} and the completeness radius relation in Equation \ref{eqn:detrad} with $a=.6$ and $b=5.23$.
With this \mainscen scenario, we find that there are \moneynumber satellites
brighter than Boo II within $417$ kpc of the Sun.  

\begin{figure}[t!]
\plotone{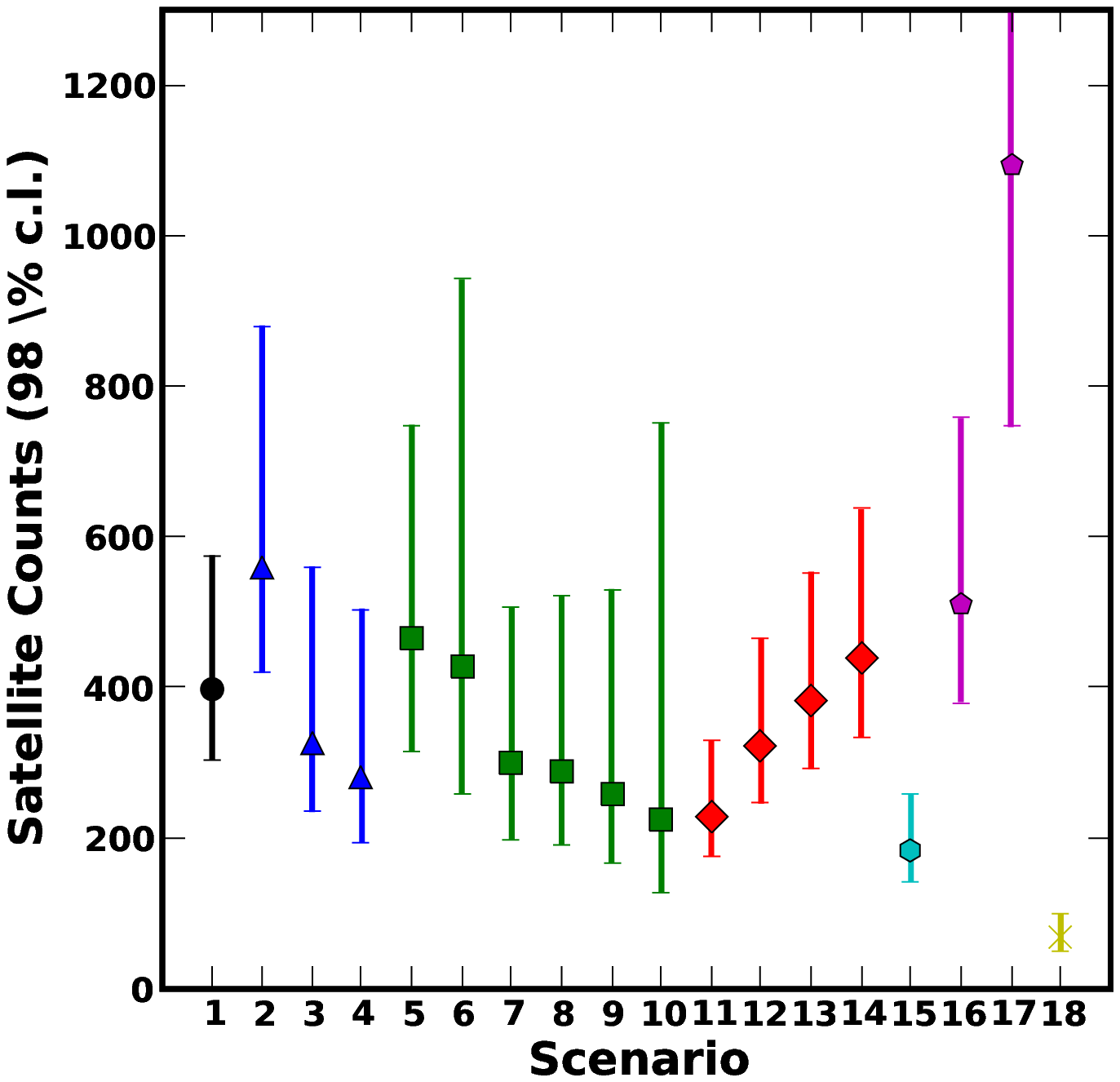}
\caption{Number of Milky Way satellites for several different scenarios as described in Table \ref{tab:res}.  Error bars reflect 98\% ranges about the median values (points).
Scenario 1 (black circle) is our \mainscen case and counts all satellites brighter than Boo II ($M_V > -2.7$) within $R_{\rm outer} = 417 kpc$.  Scenario 2-4 (blue triangles) consider inclusion or exclusion of different satellites.  Scenarios 5-10 (green squares) allow for different choices within the {\em Via Lactea} subhalo population, 11-14 (red diamonds) reflect changes in $R_{\rm outer}$, 15 (cyan hexagon) corresponds to no detection efficiency correction, and 16 and 17 (magenta pentagons) are for different completeness limit assumptions. 
Scenario 18 (yellow cross) includes only the angular (sky coverage) correction factor, i.e. no luminosity bias correction.}
\label{fig:res}
\end{figure}

We have performed the same exercise for a number of different scenarios as described in Table \ref{tab:res} and 
summarized in Figure \ref{fig:res}.  Table \ref{tab:res} assigns each of these scenarios a number (Column 1) based on 
different assumptions that go into the correction.  
Scenario 1 is our \mainscen case and counts all satellites brighter than Boo II ($M_V > -2.7$) within
$R_{\rm outer} = 417$ kpc.  
Scenario 2 counts the total number of satellites brighter than Segue 1 ($M_V > -1.5$) if we consider Segue 1 as included in the DR5 sample (Column 5).
Scenarios 3-4 exclude Boo II or Willman 1 (the faintest satellites) along with Segue 1, considering the possibility that these satellites are low luminosity owing to strong environmental effects, or are not even truly dwarf galaxies at all. Alternatively, scenario 4 may be considered the total number brighter than Coma (the 4th faintest).
Scenarios 5-10 allow for different choices within the {\em Via Lactea} subhalo population (Column 7), 
and 11-14 reflect changes in outer Milky Way radius, $R_{\rm outer}$ (Column 4).
Scenarios 16-17  allow different values for $a$ and $b$ in Equation \ref{eqn:detrad} (specified in Columns 2 and 3) for the
$R_{\rm comp}(M_V)$ relation, and 15 assumes that the detection efficiency for all
satellites is $\epsilon = 1$.   Finally, Scenario 18 makes the extreme assumption that there is no
luminosity bias and includes only a sky-coverage correction factor.
In columns 8-10,  $N_{\rm sats}$ is the number of satellites expected for a given scenario, while $N_{\rm upper}$ and $N_{\rm lower}$ are the upper and lower limits for the 98\% distribution.  The final column gives the faint-end (i.e. $ -2 \gtrsim M_V \gtrsim -7 $) slope if the luminosity function is approximated by the form $dN/dL_V \sim L^{\alpha}$. Note that some of these luminosity functions are \emph{not} consistent with a power law within the anisotropy error bars, but we include the best fits for completeness (see \S \ref{sec:conc} for further discussion).  
 
From Figure \ref{fig:res} , for all of the cases that adopt the \mainscen detection limits (scenarios 1-15), we may expect
as many as  $\sim 500$ satellites within $\sim 400$ kpc.
Even the most conservative completeness scenarios  (4,10,11, and 14)
 suggest that $\sim 300$ satellites may exist within the Milky Way's virial radius.  
Scenario 17, which relies on a less conservative but not unreasonable detection limit,  suggests that
 there may be more than $\sim 1000$ Milky Way satellites waiting to be discovered.
We now turn to a discussion of the prospects for this exciting possibility.

\begin{deluxetable*}{ccccccccccc}
\tablecolumns{11}
\tablecaption{The last four columns provide median, upper, and lower estimates (98 \% range) for the  Milky Way satellite count within $R_{\rm outer}$, as well as the faint-end slope (as the Schechter $\alpha$, i.e. $dN/dL_V \propto L_V^{\alpha}$)  for several different scenarios.  
See text for a description. }
\tablehead{
  \colhead{Scenario} &
  \colhead{$a$} &
  \colhead{$b$} &
  \colhead{$R_{\rm outer}$ [kpc]} &
  \colhead{Excluded Satellites} &
  \colhead{Detection Eff.} &
  \colhead{Subhalos} &
  \colhead{N$_{\rm sats}$} &
  \colhead{N$_{\rm upper}$} &
  \colhead{N$_{\rm lower}$} &
  \colhead{Faint-end slope($\alpha$)}
}
 \startdata
1 & -0.600 & -0.719 & 417  & Seg1 & Yes & All & 398 & 576 & 304  & $-1.87 \pm 0.19$ \\
2 & -0.600 & -0.719 & 417 & None & Yes & All & 558 & 881 & 420  & $-1.78 \pm 0.20$ \\
3 & -0.600 & -0.719 & 417 & Seg1, BooII & Yes & All & 328 & 561 & 235 & $-1.47 \pm 0.18$ \\ 
4 & -0.600 & -0.719 & 417 & Seg1, BooII, Wil1 & Yes & All & 280 & 505 & 194  & $-1.66 \pm 0.17$ \\
5 & -0.600 & -0.719 & 417  & Seg1 & Yes  & $v_{\rm max} > 5$ & 464 & 749 & 316  & $-1.92 \pm 0.23$ \\
6 & -0.600 & -0.719 & 417  & Seg1 & Yes  & $v_{\rm max} > 7$ & 427 & 942 & 258  & $-1.90 \pm 0.32$ \\
7 & -0.600 & -0.719 & 417  & Seg1 & Yes  & $v_{\rm peak} > 9$ & 302 & 508 & 198  & $-1.90 \pm 0.42$ \\
8 & -0.600 & -0.719 & 417  & Seg1 & Yes  & $v_{\rm peak} > 10$ & 289 & 521 & 191  & $-1.76 \pm 0.29$ \\
9 & -0.600 & -0.719 & 417 & Seg1  & Yes  & $v_{\rm peak} > 14$ & 260 & 531 & 167 & $-1.72 \pm 0.32$ \\
10 & -0.600 & -0.719 & 417 & Seg1  & Yes  & 65 LBA & 224 & 752 & 128  & $-1.67 \pm 0.44$ \\
11 & -0.600 & -0.719 & 200 & Seg1 & Yes & All & 229 & 330 & 176  & $-1.70 \pm 0.18$ \\
12 & -0.600 & -0.719 & 300 & Seg1 & Yes & All & 322 & 466 & 246  & $-1.80 \pm 0.19$ \\
13 &  -0.600 & -0.719 & 389 & Seg1 & Yes & All & 382 & 554 & 292  & $-1.87 \pm 0.19$ \\
14 & -0.600 & -0.719 & 500 & Seg1 & Yes & All & 439 & 637 & 335  & $-1.89 \pm 0.20$ \\
15 & -0.600 & -0.719 & 417 &  Seg1 & No & All & 184 & 259 & 142 & $-1.65 \pm 0.19$ \\
16 & -0.684 & -0.753 & 417 & Seg1 & Yes & All & 509 & 758 & 381 & $-1.95 \pm 0.20$ \\
17 & -0.667 & -0.785 & 417 & Seg1 & Yes & All & 1093 & 2261 & 746 & $-2.15 \pm 0.26$ \\
18 & N/A & N/A & 417 & Seg1 & N/A & All & 69 & 100 & 50 & $-1.16 \pm 0.21$
\enddata
\label{tab:res}
\end{deluxetable*}

\section{Prospects for Discovery}
\label{sec:future}
Upcoming large-area sky surveys such as LSST, DES, PanSTARRS, and SkyMapper \citep{lsst,des,panstarrs,skymapper} will survey more sky and provide deeper maps of the Galactic
environment than ever before possible.  In this section we provide a first rough estimate for the number
and type of satellite galaxies that one may expect to discover with these surveys.

\begin{figure}[t!]
\plotone{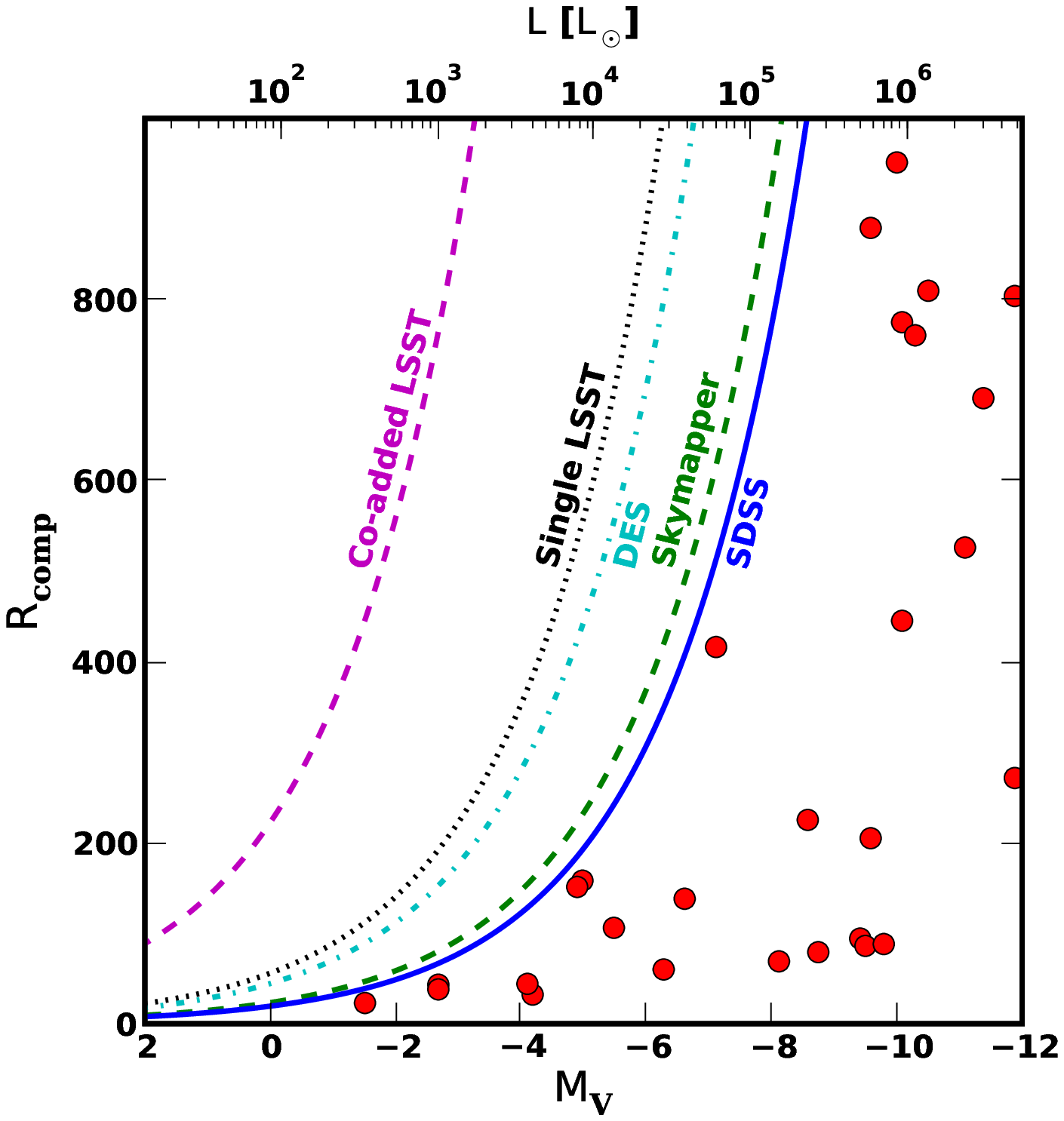}
\caption{Maximum radius for detection of dSphs as a function of galaxy absolute magnitude for DR5 (assumed limiting r-band magnitude of 22.2) compared to a single exposure of LSST (24.5), co-added full LSST lifetime exposures (27.5), DES or one exposure from PanSTARRS (both 24), and the SkyMapper and associated Missing Satellites Survey (22.6). The data points are SDSS and classical satellites, as well as Local Group field galaxies.}
\label{fig:futuresurveys}
\end{figure}

While it is impossible to truly ascertain the detection limits without detailed modelling and consideration of the sources of contamination for the magnitudes and colors these surveys will probe, the simplest approximation is to assume that all the characteristics of detectability are the same as those for SDSS aside from a deeper limiting magnitude.     If we do this, we can estimate the corresponding completeness radius for each survey
by the difference between the limiting $5-\sigma$ $r$-band point source magnitude
for Sloan ($M_{\rm SDSS} = 22.2$) and the corresponding limit for the new survey:
\begin{equation}
\frac{R_{\rm comp}^{\rm lim}}{R_{\rm comp}^{\rm SDSS}} = 10^{(M_{\rm lim}-M_{\rm SDSS})/5}.
\end{equation}
Figure \ref{fig:futuresurveys} shows the results of this exercise  for several planned surveys.  According to this estimate, LSST, with a co-added limit of $M_{\rm LSST} = 27.5$, should be able to detect objects as faint as the faintest known galaxies out to the Milky Way virial radius.  Moreover, LSST will be able to 
detect objects as faint as $L \sim 100 L_\odot$ (if they exist) out to distances of $\sim 200$ kpc.  

Examining the prospects for dwarf satellite detection with LSST in more detail, 
in Figure \ref{fig:LSSTlfs} we plot the number of satellites that LSST would detect in $4 \pi$ sterradians assuming that
the true satellite distribution follows that expected from our \mainscen correction presented above.  
Remarkably, even with single-exposure coverage (to 24.5), LSST will be able to discover a sizable fraction of the satellites in the Southern sky, $> 100$ galaxies by this estimate.  Moreover, the co-added data would reveal all $\sim 180$ dwarf galaxies brighter than Boo II (and perhaps even more if they exist at lower
luminosities).  A survey such as this will be quite important for testing galaxy formation models, as even the shape of the satellite luminosity function is quite poorly constrained.  In Figure \ref{fig:LSSTlfs},  the Skymapper line appears close to a power law over the range shown, while the ``actual" luminosity function is certainly not.  

Finally, in Table \ref{tab:futurepreds}, we present predictions for the number of satellites that will be found in each of a series of upcoming large-area surveys, as well an example smaller-scale survey, RCS-2~\citep{rcs}. Given their limiting magnitudes and their fractional sky coverage and assuming the \mainscen scenario outlined above,  we can essentially perform the inverse of the correction described in Section \ref{sec:corr}, and determine the number of satellites the survey will detect (as done in Figure \ref{fig:LSSTlfs} for LSST).  Note that this still ignores more complicated issues such as background galaxy contamination and problems with stellar crowding, which are potentially serious complications, particularly for the faintest of satellites.  Hence, these estimates are upper limits, but to first order they are a representative number of the possibly detectable satellites given the magnitude limits and sky coverage.  Some of the footprints of these surveys will overlap, however, which will improve the odds of detection, but reduce the likely number of new detections for later surveys (this effect is not included in the counts in Table \ref{tab:futurepreds}).

\begin{deluxetable}{cccc}
\tablecolumns{4}
\tablecaption{Predicted number of detectable satellites in a series of upcoming surveys (ignoring crowding/confusion effects, which will reduce the number of detections of faint objects).}
\tablehead{
  \colhead{Survey Name} &
  \colhead{Area (deg$^2$)} &
  \colhead{Limiting $r$ [mag]} &
  \colhead{$N_{\rm sats}$} 
}
 \startdata
RCS-2 & 1000 & 24.8 & 3-6 \\
DES & 5000 & 24 & 19-37 \\
Skymapper & 20000 & 22.6 & 42-79 \\
PanSTARRS 1 & 30000 & 22.7 & 61-118 \\
LSST 1-exp & 20000 & 24.5 & 93-179 \\
LSST combined & 20000 & 27.5 & 145-283 
\enddata
\label{tab:futurepreds}
\end{deluxetable}

\section{Discussion}
\label{sec:disc}

 \subsection{Caveats}
  
While our general expectation that there should be many more satellites seems to be quite solid, 
there are several important caveats that limit our ability to firmly quote a number or to discuss the expected overall shape of the
luminosity function.

\begin{enumerate}

\item Detection Limits.   The simplifying approximation that Equation \ref{eqn:complim} describes the complexities of satellite detection cannot be correct in detail -- we expect a dependence on properties such as Galactic latitude and color.   However, we are leaving these issues aside for this first-order correction.  Moreover, the detection limits from \citet{kop07} apply only to objects with surface brightness $\mu_V \lesssim 30$; if dSph's more diffuse than this exist, they will have been missed by SDSS.  In this sense, our results are conservative, as even more very low surface brightness dwarf galaxies may be discovered by new, deeper surveys.

\item Input Assumption.  The assumption that the satellite population tracks the full underlying subhalo population is a major assumption of this correction.  As Figure \ref{fig:raddist}  shows, this assumption is consistent with the current data,  and modest cuts on  $v_{peak}$ improve the situation further.   Given the size of the error bars, the existence of a discrepancy simply cannot be resolved without surveys that have better completeness limits.  There are certainly reasons to expect that the satellite population {\em will not} track the subhalo population in an unbiased way -- for example, if tidal forces play a role in making ultra-faint dSph galaxies.  However, there are also reasons to suspect that there is no such bias.   The dark matter subhalo masses of the known Milky Way dwarfs are approximately the same over a $\sim 4 $ orders-of-magnitude spread in luminosity  \citep[Strigari et al. 2008, in prep.;][]{strigari08}, suggesting that the luminosity that any subhalo obtains is quite stochastic.  Whether or not our input assumption about galaxies tracing subhalos is correct, future surveys will provide a means to test it, and thereby provide an important constraint on the formation processes of these extreme galaxies.

\item Subhalo Distribution.  Our correction relies on the \emph{Via Lactea} subhalo population and its distribution with radius and angle.  While this provides a well-motivated starting point for this correction,  \emph{Via Lactea} is only a single realization of a particular mass halo and of a particular set of cosmological parameters (WMAP 3 yr), and hence cannot necessarily be considered representative of a typical halo in $\Lambda$CDM.  Semi analytical models suggest that the cosmic variance in radial distributions should not be very large, as long as we consider halos that are small enough that the mass-function is well populated \citep{ZB03}.  However,  a suite of \emph{Via Lactea}-type simulations will be necessary to produce a distribution of corrections to statistically average over before this correction can be considered fully numerically robust.  As discussed in \S \ref{sec:intro}, the uncertainty in the properties of the Milky Way halo also mean that using \emph{Via Lactea} to accurately model the Milky Way is subject to those same uncertainties.  This uncertainty is alleviated in this correction because only the fraction distribution matters -- the absolute normalization is determined by the \emph{observed} satellites rather than the subhalo counts.  But if it is true that the radial distribution of subhalos tracks the global dark matter distribution, as suggested  by~\citet{die07ms} and Figure \ref{fig:nvsr}, this uncertainty could be important if the inner portions of the radial distribution change substantially with cosmological parameters or halo properties. 

Note that while the mass of \emph{Via Lactea} is quite close to that obtained from $\Lambda$CDM-based mass models of the Milky Way \citep[e.g.][]{kly02mwmass,madau081e8ell},  the mass of the Milky Way halo is constrained by observations only at the factor of $\sim 2$ level.   The most relevant uncertainty associated with the possible mass difference between the Milky Way and \emph{Via Lactea} is the difference associated with the uncertain radial scaling. Since $R_{200} \propto M_{200}^{1/3}$, a factor of $\sim 2$ mass uncertainty corresponds to a  $\sim \pm 25 \%$  uncertainty in virial radius. Fortunately, our corrections depend primarily on the relative radial distribution of satellites, not on their absolute masses, and the choice of exactly what radius within which galaxies are considered ``satellites'' is somewhat arbitrary, anyway.   Moreover,  the expected subhalo angular anisotropy (see \S \ref{sec:corr}) is more important for our corrections than the uncertainty associated in the Milky Way halo mass or precisely how the outer radius is defined.

\item Input Luminosities/Satellites.  Most of our corrections come from the faintest, nearby satellites, as they have very small $R_{\rm{comp}}$ values compared to $R_{\rm{outer}}$.  Errors in the magnitudes of these faintest of satellites will result in significant changes in the derived luminosity function.  These errors are not propagated for this correction, but revisions to faint satellite magnitudes have the potential to significantly alter our estimates.  Furthermore, the final count is highly sensitive to Poisson statistics in the innermost regions of the Milky Way, as well as any physical effect that biases the radial distribution of the faintest of the satellites (as discussed above). 	
\end{enumerate}

\subsection{Implications}
\label{sec:impl}
 Keeping in mind the caveats discussed above, it is worth considering the potential implications of a Milky Way halo filled with hundreds of satellite galaxies.  For the sake of discussion, we adopt $N_{\rm sat} \simeq 400$, as obtained in our fiducial estimate.

First, $N_{\rm sat} \simeq 400$ can be used to determine a potential characteristic velocity scale in the subhalo distribution. Under the
(extreme) assumption that only halos larger than a given velocity host satellites,   Figure \ref{fig:nvsvmax}  implies that $N \simeq 400$ corresponds to a
 circular velocity $v_{\rm max} \gtrsim 7$ \kms or a historical maximum of
   $v_{peak} \gtrsim 12$ \kms.  We note that these cutoffs correspond closely to those adopted in our corrections (Table \ref{tab:res}) and that there was no reason that they had to match.  If we iteratively apply the correction in order to obtain
 perfect self-consistently between the corrected count and the input total, we find $v_{peak} > 14$ \kms.    Interestingly, this scale is quite close to the threshold where gas may be boiled out of halos via photoevaporation \citep{BL99}.  We note however that the subhalo abundance at fixed mass or $v_{\rm max}$  may vary perhaps at as much as the factor of $\sim 2$ level from halo to halo \citep{ZB03}.  Therefore the ``cosmic variance error" will affect our ability to determine a characteristic subhalo mass based on counts (but does not affect the correction itself, which depends only on the radial distribution, not the total counts of subhalos).  Recently, \citet{diemand08} have published results from the \emph{Via Lactea II} halo, which has a factor of $\sim 1.7$ times more subhalos at fixed $v_{\rm max}$ than the \emph{Via Lactea} halo we analyze here.  If we take the velocity function in \citet{diemand08} Figure 3, $N_{\rm sat}$ would correspond to $v_{\rm max} \gtrsim 9$ \kms.

While the maximum circular velocity of a subhalo is a useful measure of its potential well depth, it is very difficult to measure $v_{\rm max}$ directly from dwarf satellite stellar velocities \citep{str07redef}. The best-determined dark halo observable is the integrated mass within a fixed radius within the stellar distribution \citep{SBK}. While the observed half-light radii vary from tens to hundreds of parsecs for the dwarf satellites, all dwarf satellites are found to have a common mass scale of $\sim 10^7 M_\odot$ within a fixed radius of 300 pc within their respective centers (Strigari et al 2008 in prep.) and to a similar extent a common mass of $10^6$ within 100 pc (Strigari et al 2008). Although the masses within these scales are difficult to resolve with the {\emph Via Lactea} simulation we consider in this paper, this mass scale will be well-resolved in \emph{Via Lactea II} \citep{diemand08} and forthcoming simulations.  In the future, a statistical sample of highly resolved subhalos will allow for a robust comparison between the dwarf satellite mass function and the subhalo population.  This in turn will allow corrections of the sort presented in this paper on various sub populations of subhalos and galaxies, providing a much more stringent consistency check between the mass function and luminosity function in $\Lambda$CDM-based models.  It is important to note, however, that the observed stellar kinematics of the satellites does set stringent limits on their host halo $v_{\rm max}$ values.  These limits are consistent with the results presented here in that any of the reasonable sub-populations discussed above (e.g. $v_{\rm max} \gtrsim 7-10$ \kms) are not excluded by the current data \citep{str07redef,strigari08}.   We note that strong CDM priors suggest somewhat \emph{larger} $v_{\rm max}$ values \citep[$\gtrsim 15$\kms --][]{str07redef,strigari08,pen08}, which would (if anything) \emph{underpredict} the observed visible satellite counts, according to our estimates. 

As we have shown,  we expect the discovery of many more dwarfs to occur with planned surveys like LSST, DES, PanSTARRS, and SkyMapper.  If so, it will provide important constraints on galaxy formation models, which, at present, are only poorly constrained by the present data. The ability to detect galaxies
as faint as $L \sim 100 L_\odot$ provides an opportunity to discover if there is a low-luminosity threshold in galaxy formation, and to use these faint galaxies as laboratories to study galaxy formation in the extreme.  The planned surveys will also provide an important  measurement of the radial distribution of faint satellites.  This will help test our predictions, but more importantly provide constraints on more rigorous models aimed at understanding why and how low-mass galaxies are so inefficient at converting their baryons into stars. 

Another important direction to consider is searches for similar satellites around M31.  New satellites are rapidly being discovered in deep surveys of its environs \citep[e.g.][]{mcconn08newands}.  While there are indications of substantial differences in the M31 satellites and their distributions compared to the Milky Way \citep{mcconn06}, such comparisons are complicated by the fact that it is impossible to detect the ultra-faint satellites that comprise most of our corrected satellite counts at the distance of M31.   with the much larger data samples that will be available with future deep surveys, it may be easier to compare the true luminosity function and distributions of Milky Way satellites to M31 and hence better understand the histories of both the galaxies, as well as better constrain how dSph's form in a wider $\Lambda$CDM context.

Finally, if LSST and other surveys do discover the (full-sky) equivalent of
$\sim 400$ or even $\sim 1000$ satellites, and appropriate kinematic follow-up with 30m-class telescopes like the Thirty Meter Telescope (TMT) confirms  that these objects were indeed dark-matter dominated, then
 it will provide a unique and powerful means to constrain the particle nature
of dark matter.   As discussed in the introduction, the mass function of dark matter subhalos is expected to rise steadily to 
small masses as $\sim 1/M$  \citep{kl99ms,die06vl} and the only scale that is expected to break this rise is the 
cutoff scale in the clustering of dark matter.  When the MSP was originally formulated,
scenarios like warm dark matter (WDM) were suggested as a means of ``erasing" all but the $\sim 10$ most massive
subhalos per galaxy by  truncating the power at $\sim 10^8 \Msun$ scales.  If $\sim 1000$ satellites were discovered,
then the same idea could be used to provide a limit on the small-scale clustering characteristics of the dark matter
particle.  As a rough approximation, $N \simeq 1000$ subhalos corresponds to a minimum mass subhalo in
\emph{Via Lactea} of $M \simeq 10^7 \Msun$ \citep{die06vl}, or an 
original mass (before infall) of $M_i \simeq 3 \times 10^7 \Msun$ \citep{die07ms}.
 If we associate this $z=0$ subhalo mass with a limiting free-streaming mass, then we obtain the bound
 $m_\nu \gtrsim 10$ keV on the sterile neutrino \citep{abazajian06}.  This limit is competitive with the best constraints possible with the Ly$\alpha$ forest and is not subject to the uncertainties of baryon physics.  Of course, WDM simulations
 will be required in order to convincingly make a link between satellite counts and the small-scale power-spectrum,
 but these simulations are certainly viable within the time frame of LSST.

\begin{figure}[t!]
\plotone{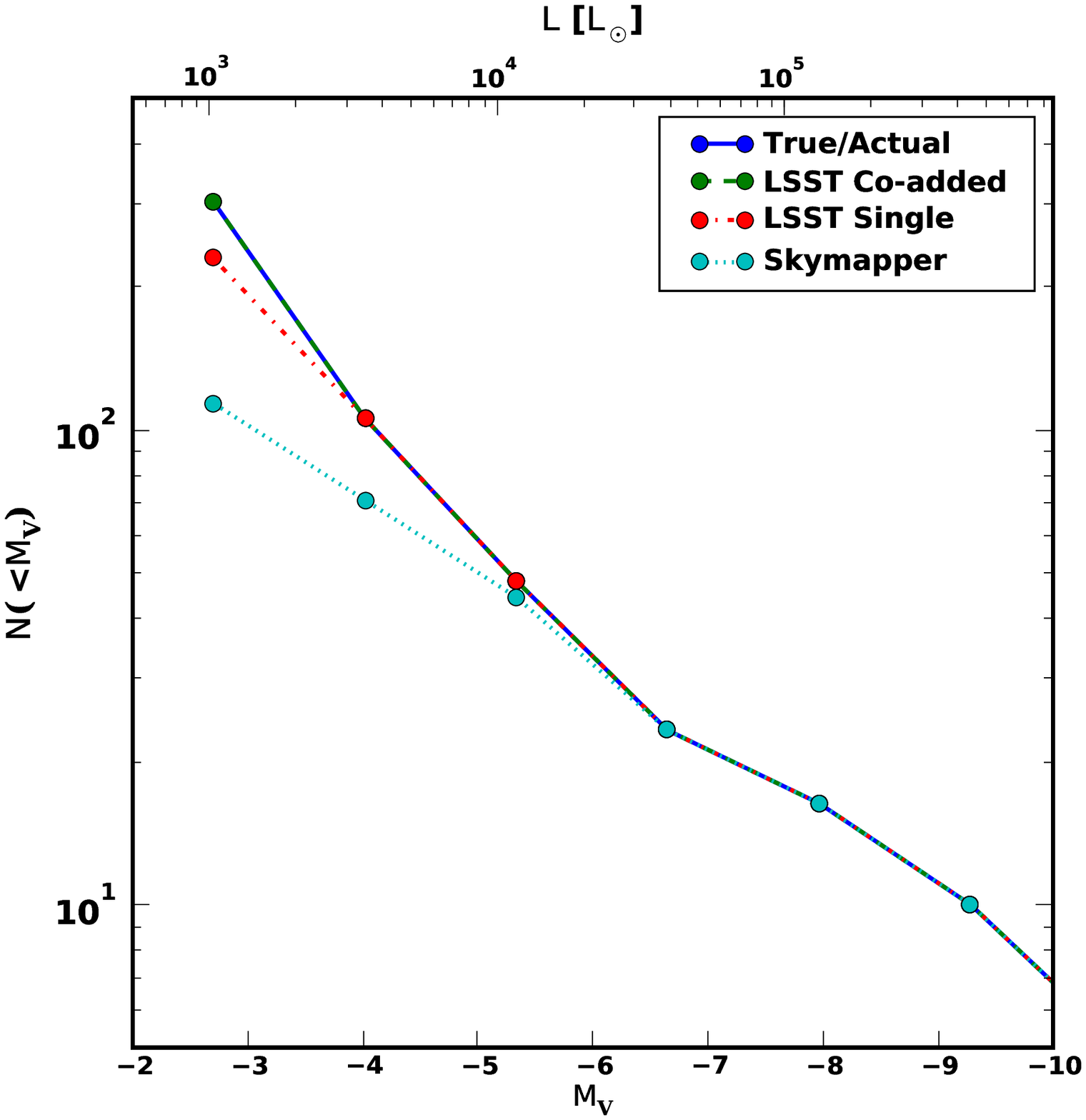}
\caption{Expected luminosity functions for LSST per $4 \pi$ sterradians.  For the single exposure, the adopted limiting magnitude is $r_{\rm lim}=24.5$, while for the co-added case, $r_{\rm lim}=27.5$  The Skymapper curve assumes a hypothetical full sky coverage survey of the same limiting magnitude as Skymapper ($r_{\rm lim}=22.6$).}
\label{fig:LSSTlfs}
\end{figure}

\section{Conclusions}
\label{sec:conc}
The goal of this paper has been to provide reasonable, cosmologically-motivated corrections to the observed luminosity function.   Our primary aim is to motivate future searches for faint dwarf galaxies and to explore the status of the missing satellites problem in light of the most recent discoveries.   By combing completeness limits for the SDSS \citep{kop07} with the spatial distribution of subhalos in \emph{Via Lactea}, we have shown that there are likely between $\sim 300$ and $\sim 600$ satellites brighter than Boo II within the Milky Way virial radius (Figure \ref{fig:moneyplot}),  and that the total count may be as large as $\sim 2000$, depending on assumptions (Table \ref{tab:res}, Figure \ref{fig:res}).
We also showed that the observed satellites are indeed consistent with tracing the radial distribution of subhalos, provided completeness limits are taken into account (Figure \ref{fig:raddist} and Table \ref{tab:radKS}).  Moreover, we argued that future large sky surveys like LSST, DES, PanSTARRS, and 
SkyMapper should be able to see these satellites if they do exist, and thereby provide unprecedented constraints on the nature of galaxy formation in tiny halos.

While this correction predicts that nearly all of the undetected satellite are faint ($M_V>-7$) and consequently have low surface brightness, it is important to note that it  is not clear how this maps onto satellites' $v_{\rm max}$.  While a possible test of this correction's result lies in selecting sub-samples of the observed satellite population, $v_{\rm max}$ is difficult to constrain in the known satellites \citep{str07redef}, and hence this exercise is suspect until simulations can resolve subhalos well enough to compare to observables such as the integrated mass within 300 pc (see \S \ref{sec:impl}).

There are two major points to take away from this correction:
\begin{itemize}

\item  As it was first formulated, the MSP referred to the mismatch between the then $\sim 10$ known dwarf satellite galaxies of the Milky Way and Andromeda, and the expected count of $\sim 100-500$ subhalos with $v_{\rm max} \ge 10$ \kms \citep{kl99ms,moo99ms}.    Our results suggest that the recent discoveries of ultra-faint dwarfs about the Milky Way are consistent with a total population of $\sim 500$ satellites, once we take into account the completeness limits of the SDSS.  In this sense, the primary worries associated with the MSP in CDM are alleviated.  Nonetheless, it is critical that searches for these faint galaxies be undertaken, as the assumptions of this correction must be tested.

\item  The shape of the faint end of the satellite luminosity function is not yet constrained well enough to deeply understand the theoretical implications.  There exists yet a large parameter space in galaxy formation theory that will fall inside the error bars of Figure \ref{fig:moneyplot}, and an even larger parameter space of models that are viable if our caveats and scenario possibilities are considered.  Our results strongly suggest that the luminosity function continues to rise to the faint end, with a faint end slope in our fiducial scenario given by $dN/dM_V = 10^{(0.35\pm0.08) M_V + 3.42 \pm 0.35}$, or $dN/dL_V \propto L^\alpha$, with $\alpha=-1.9 \pm 0.2$.  This is substantially steeper than the result of \citet{kop07}, possibly the result of using a $\Lambda$CDM-motivated subhalo distribution instead of the analytic profiles used in that paper.  But caution is advised in reading anything into the details of the shape, as nearly anything could be hiding within the faintest few bins when all the scenarios are considered.  This is apparent from Figure \ref{fig:LSSTlfs}, where the Skymapper line appears as a power law over the range shown, while the actual luminosity function is certainly not.  Furthermore, depending on which scenario is used, the slope ($\alpha$) can vary anywhere from -1.16 to -2.15 (see last column of Table \ref{tab:res}).  

\end{itemize}

Fortunately, future deep large sky surveys will detect very faint satellites out to much larger distances and hence firmly observe the complete luminosity function out beyond the Milky Way virial radius (see Section \ref{sec:future}).  With these data, it will be possible to provide stringent limits both on cosmology and galaxy formation scenarios (see Section \ref{sec:impl}).  Nonetheless, the current data are not deep enough, and until the new survey data are available, there will be no way to put the spectre of the MSP completely to rest.

\section{Acknowledgements}
We wish to acknowledge the creators of the \emph{Via Lactea} Simulation and the public data release thereof (\url{http://www.ucolick.org/\~{}diemand/vl/data.html}).  We thank Manoj Kaplinghat, Betsy Barton, Anatoly Klypin, J{\"u}rg Diemand, Piero Madau, Ben Moore, Andrew Zentner, Nicolas Martin, and Alan McConnachie for helpful conversations and suggestions, as well as Brant Robertson for conversations and bringing the RCS-2 survey to our attention.  We would also like to thank the anonymous referee for helpful comments and suggestions.  This work was supported by the Center for Cosmology at UC Irvine.

\bibliography{ms}{}
\bibliographystyle{hapj} 

\end{document}